\documentclass[12pt]{article} 
\input{epsf.tex}
\usepackage{epsfig,ldd_art }
\newcommand{\One}{1\kern-4.5pt1}
\newcommand{\lapprox}{\raisebox{-0.5ex}{$\ 
\stackrel{\textstyle<}{\textstyle\sim}\ $}}
\newcommand{\gapprox}{\raisebox{-0.5ex}{$\ 
\stackrel{\textstyle>}{\textstyle\sim}\ $}}
\begin{document}
\vspace{-20pt}

\addtolength{\baselineskip}{0.20\baselineskip}

\hfill SWAT/01/287

\begin{center}
{\Large{\bf The Phase Diagram of QCD}}
\medskip

\centerline{\bf Simon Hands }

\centerline{\sl Department of Physics, University of Wales Swansea,}
\centerline{\sl Singleton Park, Swansea SA2 8PP, U.K.}
\end{center}

\noindent
{\narrower 
{\bf Abstract:} I use simple thermodynamic reasoning to argue that at
temperatures of order a trillion kelvin, QCD, the theory 
which describes strongly interacting particles such as protons and
neutrons under normal conditions, undergoes a phase transition to a plasma of 
more elementary constituents
called quarks and gluons. I review what is known about the plasma phase both 
from theoretical calculations and from experiments involving the collisions
of large atomic nuclei moving at relativistic speeds. Finally I consider the
behaviour of nuclear material under conditions of extreme density, and discuss
possible exotic phenomena such as quark matter and color superconductivity.}
\vfill
\begin{figure}[htb]
\begin{center}
\epsfig{file=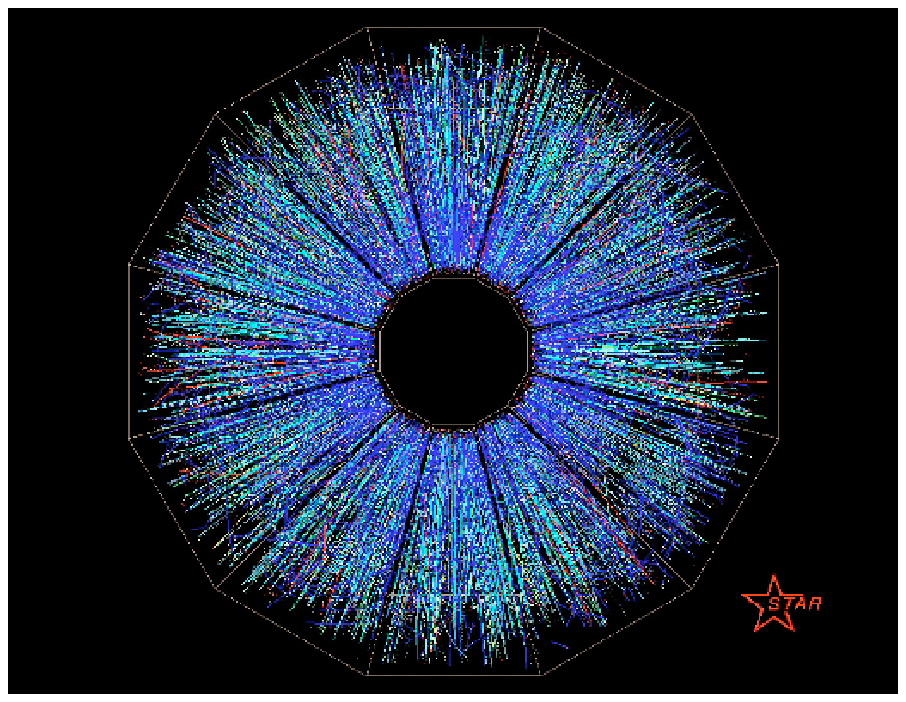, width=11cm}
\caption{Au+Au collision in the STAR detector at 
the Relativistic Heavy Ion Collider \label{fig:RHIC}}
\vskip -10pt
(courtesy Brookhaven National Laboratory)
\end{center}
\end{figure}

\newpage

\section{Introduction}

For over 25 years, our best theory of the strong interaction responsible for
binding together protons and neutrons within the atomic nucleus has been
Quantum Chromodynamics, or QCD for short. QCD treats nucleons not as
fundamental objects in their own right, but as composite states of
size roughly 1fm~\footnote{fm stands for {\sl femtometre\/}, 
also known as a {\sl fermi\/}.
$1\mbox{fm}=10^{-15}\mbox{m}$.} made from more
elementary particles called {\sl quarks\/}. Quarks have spin-${1\over2}$ and 
come in several varieties, or
{\sl flavors\/}; the two lightest quarks are the up quark $u$, which carries 
an electric charge $+{2\over3}e$, where $e$ is the absolute value of the
electron charge, and the down quark $d$ which has charge $-{1\over3}e$. 
Table~\ref{tab:quarks} summarises the important properties of the known quarks.
Note that we specify both ``constituent'' and ``current'' quark masses, 
a distinction to
be explained below.
Constituent and current masses can differ considerably for the 
lighter flavors $u$, $d$, $s$ -- 
the origin of the $O(10^4)$ factor
between 
heaviest and lightest current masses is as yet not understood.
With an eye to the charge and constituent mass
assignments of Table \ref{tab:quarks}, 
we can identify a proton 
as a $uud$ bound state with net charge $+e$ and the neutron
as $udd$ with net charge zero. 

\begin{table}[htb]
\setlength{\tabcolsep}{1.5pc}
\caption{Summary of properties of the known quarks}
\label{tab:quarks}
\begin{tabular*}{\textwidth}{@{}l@{\extracolsep{\fill}}llllll}
\hline
quark flavor && symbol & charge  & constituent mass & current mass \\
&& & $Q/e$ & $\Sigma$ (MeV/$c^2$) & $m$ (MeV/$c^2$)\\
\hline
down && $d$ & -${1\over3}$ & $\sim350$ & $\sim7$  \\
up && $u$ & +${2\over3}$ & $\sim350$ & $\sim3$  \\
strange && $s$ & -${1\over3}$ & $\sim550$ & $\sim140$  \\
charm && $c$ & +${2\over3}$ & $\sim1800$ & $\sim1800$  \\
bottom && $b$ & -${1\over3}$ & $\sim4.2\times10^3$ & $\sim4.2\times10^3$  \\
top && $t$ & +${2\over3}$ & $\sim170\times10^3$ & $\sim170\times10^3$  \\
\hline
\end{tabular*}
\end{table}
The experimental support for this picture is
two-fold. Firstly, the quark model provides a very natural explanation for the 
multiplicity and pattern 
of strongly-interacting particles (collectively known as 
{\sl hadrons\/}) 
briefly formed in high-energy particle collisions \cite{quarks}. 
Hadrons can be classified as either three-quark
$qqq$ states called {\sl
baryons\/}, such as nucleons, or quark -- anti-quark $q\bar q$ states called
{\sl mesons\/}, the lightest example being the pi-meson or pion, which 
occurs in three charge states $\pi^{\pm}$ and $\pi^0$, all with
roughly one-seventh the mass of the proton. 
Apart from nucleons and pions,
most hadrons decay via the strong interaction and have typical lifetimes 
of $O(10^{-23}\mbox{s})$. 
Their variety can
be accounted for by differing combinations of quark spin and flavor, 
as well as various radial excitation
states. 
Secondly, the results of high energy inelastic
scattering experiments of 
electrons off nucleons are consistent with the presence of pointlike
spin-${1\over2}$ constituents called {\sl partons\/}, just as high-angle
Rutherford scattering of $\alpha$-particles off atoms 
demonstrates the existence of the nucleus. 
The partons appear to able to move freely
within the nucleon volume \cite{partons},
and it is natural to identify them with quarks.
In still higher energy $e^+e^-$ experiments rather 
well-collimated sprays or {\sl jets\/} of particles,
all with similar momenta, 
emerge from the event region. These are most naturally explained by
a single progenitor quark scattered in the original high-energy collision,
which subsequently decays into
a profusion of lower energy hadrons. Paradoxically, however, although 
quark constituents provide an economical explanation of strong
interaction phenomena, no experiment has ever revealed evidence for an isolated
quark, for instance via the observation of a fractionally-charged object on
a Millikan oil-drop.

\begin{figure}[htb]
\begin{center}
\epsfig{file=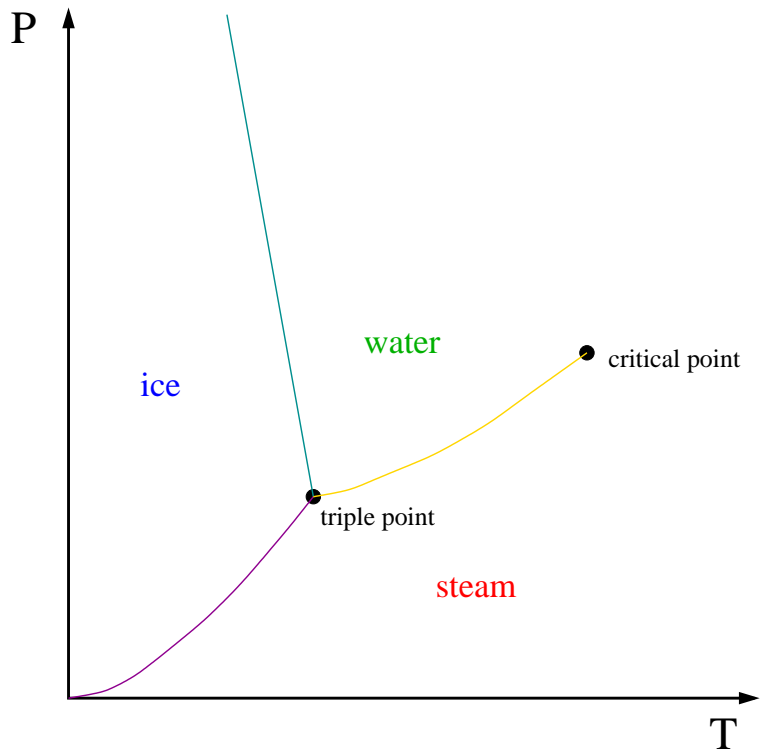, width=10cm}
\caption{Phase diagram for H$_2$O (not to scale)\label{fig:H2O}}
\end{center}
\end{figure}
In this article I wish to explore the consequences of the quark picture for the
thermodynamics of strongly-interacting matter, in other words, its behaviour as
conditions such as temperature and density are varied. Thermodynamical
information is often presented in the form of a {\sl phase diagram\/}, 
in which the
different manifestations or phases of a substance occupy different regions of a
plot whose axes are calibrated in terms of the external conditions or {\sl
control parameters\/}. The most familiar example, of course, is that of H$_2$O
shown in Fig.~\ref{fig:H2O};
the control parameters are temperature $T$ and pressure $P$, and the 
three regions correspond to the three phases of ice, water and steam. The lines
mark the various coexistence curves $P(T)$ where two phases are in equilibrium;
a phase transition such as melting or boiling
is observed when moving along a path in the $(T,P)$ plane
which intersects such curves.
Two special points in the diagram 
are the {\sl triple point\/} ($T_{tr}=273.16$K,
$P_{tr}=600\mbox{Nm}^{-2}$) where all three phases coexist, and the {\sl
critical point} ($T_c=647$K, $P_c=2.21\times10^7\mbox{Nm}^{-2}$), where the
meniscus separating liquid from vapour disappears, and the two fluid phases
become indistinguishable. For $T<T_c$ the transition between liquid and vapour
is {\sl first-order\/}, implying discontinuities $\Delta S$ and $\Delta V$ in
entropy and volume respectively, and a non-vanishing latent 
heat and interface tension. This classification follows because
entropy and volume are both first derivatives of the Gibbs free energy $G(T,P)$:
\begin{equation}
S=-{{\partial G}\over{\partial T}}\biggr\vert_P\;\;;\;\;
V= {{\partial G}\over{\partial P}}\biggr\vert_T.
\end{equation}
At the critical point the transition becomes {\sl second order\/}, which means 
that
singularities instead occur in specific heat $C_P$
and isothermal compressibility $\kappa_T$ of the fluid, 
which are related to second derivatives of the free energy: 
\begin{equation}
C_P=-T{{\partial^2 G}\over{\partial T^2}}\biggr\vert_P\;\;;\;\;
\kappa_T=-{1\over V}{{\partial^2 G}\over{\partial P^2}}\biggr\vert_T.
\label{eq:2nd}
\end{equation}
In fact, each of the quantities in eqn.(\ref{eq:2nd}) diverges at the critical
point. Another interesting phenomenon is {\sl critical
opalescence\/}: the size of droplets of the liquid phase 
within the vapour (or {\it vice
versa\/}) becomes 
comparable with the wavelength of visible light, implying large
optical path differences between adjacent parallel 
rays of light and hence strong
scattering -- the system thus becomes opaque near the critical point. 
Just beyond the
critical point, thermodynamic observables still vary very rapidly as one moves
about the $(T,P)$ plane due to the
large values of $C_P$ and $\kappa_T$; this is known as a {\sl crossover\/}
region.

\begin{figure}[htb]
\begin{center}
\epsfig{file=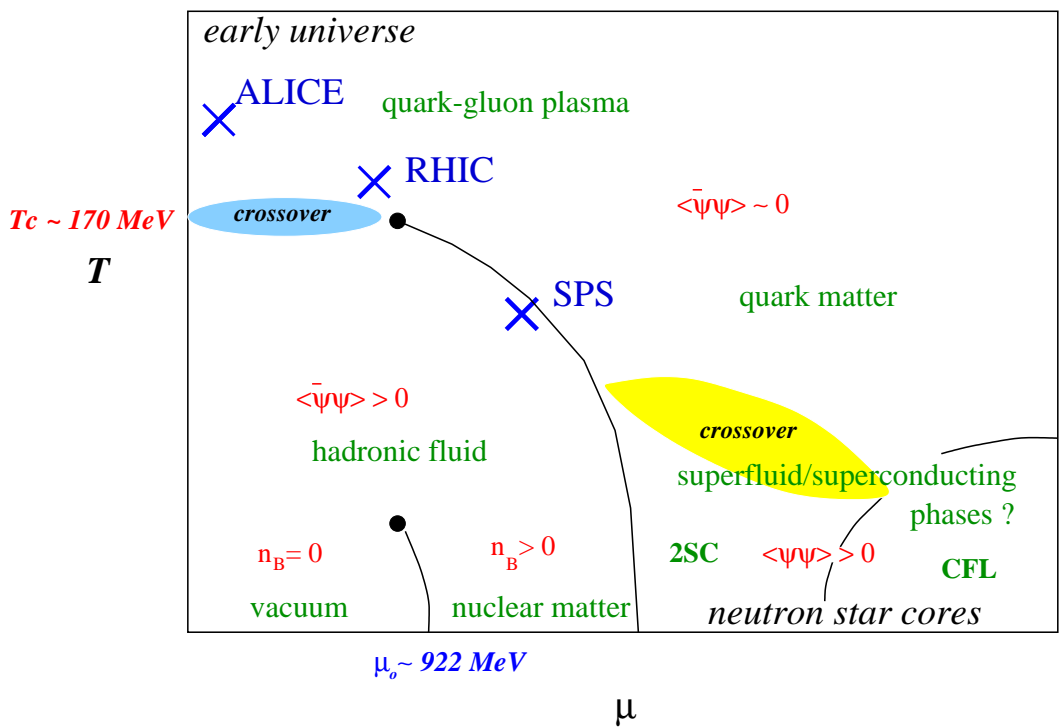, width=14cm}
\caption{Proposed phase diagram for QCD. SPS, RHIC and ALICE are the names
of relativistic heavy-ion collision experiments 
described in section~\ref{sec:rhic}. 2SC and CFL refer
to the diquark condensates defined in eqns.~(\ref{eq:2SC},\ref{eq:CFL}).
\label{fig:qcdphase}}
\end{center}
\end{figure}
Fig.~\ref{fig:qcdphase} 
shows a proposed phase diagram for QCD. The names of the various phases
are shown in green, and the environment in which they might be found in black.
Phase coexistence
lines are shown as solid lines, critical points as filled circles,
and crossovers by shaded regions. The control parameters are temperature $T$ 
and {\sl baryon chemical potential\/} $\mu$, 
with which it is probably appropriate to
(re-)familiarise ourselves. In particle physics reactions a $qqq$ baryon is
always
created or destroyed pairwise with a $\bar q\bar q\bar q$ anti-baryon. 
There is no process within QCD which can change the
number of baryons $N_B$ minus the number of anti-baryons $N_{\bar B}$; 
in other words we can
identify a conserved quantum number $B=N_B-N_{\bar B}$ 
called {\sl baryon number\/}. Quarks
and anti-quarks carry $B=\pm{1\over3}$ respectively. Now, for systems in which 
baryon number is allowed to vary, the most convenient themodynamic potential
to consider is the {\sl grand potential\/} $\Omega(T,V,\mu)=E-TS-\mu B$. 
Thermodynamic 
equilibrium is reached when $\Omega$ is minimised, and for a system in
equilibrium we recognise $\mu$ as the increase in $E$ whenever
$B$ increases
by one. When systems are analysed using the {\sl grand canonical ensemble\/} 
$\mu$ is kept as a control parameter, and the baryon density $n_B=B/V$ is a 
derived quantity whose value depends on the details of the equation of state
$n_B=n_B(T,\mu)$.

It is worth making two further observations. 
Firstly, baryons are fermions and are
hence governed by the Pauli Exclusion Principle, implying that no two identical
baryons can share the same quantum state. 
At $T=0$ we thus expect baryonic matter to
be {\sl degenerate\/}, implying that energy states are fully 
populated up to some
maximum energy called the Fermi energy $E_F$,
in precise analogy with condensed matter
systems such as electrons in a metal. 
For weakly interacting baryons
$\mu$ coincides with $E_F$.
Secondly, QCD is a relativistic quantum
field theory, which means that in contrast with electronic systems 
the particles' rest-energy should be taken into account; 
we thus have $\mu=\mu_{NR}+m_0c^2$, where
$\mu_{NR}$ is the non-relativistic chemical potential.

In the rest of this article I shall attempt to explain what is known about the 
phase diagram, giving quantitative details wherever possible. 
In the next section
I will review QCD at zero temperature and chemical potential, paying particular
attention to the question of why isolated quarks have not been
observed, and why 
observed baryon masses are considerably greater than the $u$ and $d$ current
quark
masses shown in Table~\ref{tab:quarks}. In the bottom left-hand corner of 
the phase diagram where $T$ and $\mu$ are both small
the thermodynamic behaviour of QCD can be described 
in terms of a vapour of hadrons, 
which as we have seen are composite states of quarks
and/or anti-quarks. The
principal task for QCD theorists in this region is therefore to classify and
quantify the 
bound states; there is a sense in which this can be caricatured as 
``relativistic atomic physics''. In section \ref{sec:thermo} I will use simple
thermodynamic reasoning to argue that this state of affairs cannot persist as
$T$ is raised -- eventually there comes a point where either a transition
or a crossover occurs to a 
phase where the dominant degrees of freedom are no longer
hadrons but the quarks themselves, together with other partons called {\sl
gluons\/}. Since quarks and gluons play similar roles in QCD to the
electrons and photons of QED, this phase is often called the 
{\sl quark-gluon plasma\/}
(QGP), and QCD in the upper-left region of the diagram shares much
terminology with 
relativistic plasma physics. In section \ref{sec:rhic} I discuss where the QGP
might be observed; since the required temperatures turn out to be 
extraordinarily
large, the only two candidates are the early universe, in the first instants
following the Big Bang, and in high energy collisions between not elementary
particles, but entire atomic nuclei 
which are first stripped of their attendant
electrons and then accelerated to relativistic speeds. Such experiments are
currently being performed at the Relativistic Heavy Ion Collider (RHIC)
in Brookhaven National Laboratory, New York, and still
more energetic ones are planned at the ALICE~\footnote{{\bf A} {\bf L}arge 
{\bf I}on {\bf C}ollider {\bf E}xperiment}
detector at CERN in Geneva.
Finally, I switch attention along the $\mu$-axis, where very little is known 
beyond the onset of nuclear matter at $\mu_o\simeq922\mbox{MeV}$, given
by the nucleon rest mass minus the binding energy per nucleon, which 
is estimated from empirical models of nuclei such as the liquid drop model.
Once again, there is believed to be a phase transition at a larger value of
$\mu$ to a phase in which quarks rather than nucleons are the dominant degrees
of freedom. Such {\sl quark matter\/} may conceivably be found at the cores
of compact astrophysical objects such as neutron stars.
The nature of quark matter has recently been the focus of intense theoretical
interest; it has been speculated that Fermi surface phenomena analogous to 
the Bardeen-Cooper-Schrieffer (BCS)
instability, responsible for superconductivity in metals 
and superfluidity in liquid $^3$He at low temperatures,
may play an important role. 
In the lower-right region of the phase diagram,
therefore, QCD becomes a branch of
condensed matter physics.

\section{Vacuum QCD}
\begin{figure}[htb]
\begin{center}
\epsfig{file=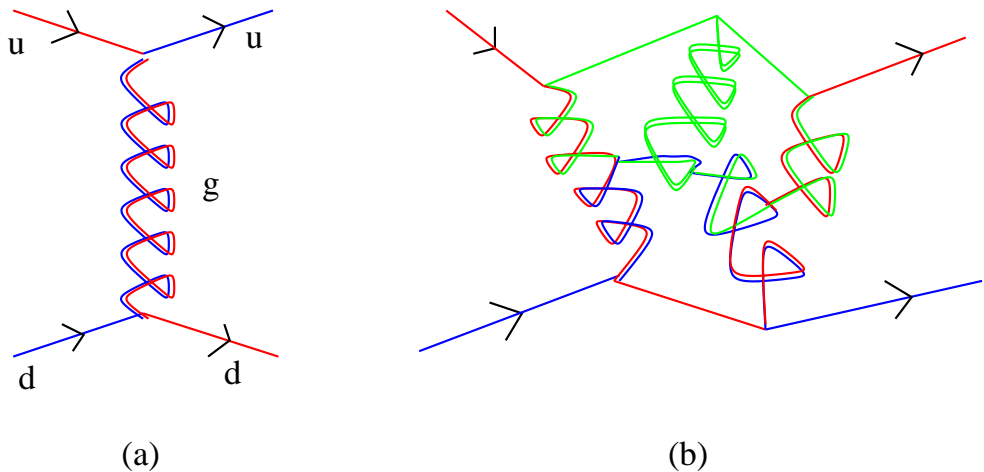, width=13cm}
\caption{Gluon exchange between quarks. For small inter-quark separation,
single gluon exchange ({\sl a\/}) dominates. For larger separations
gluon self-interactions ({\sl b\/}) are also important.\label{fig:confine}}
\end{center}
\end{figure}
The fundamental interaction between quarks in QCD arises from the exchange of
spin-1 particles called {\sl gluons\/} $g$, shown schematically in
Fig.~\ref{fig:confine}. Gluons are present inside hadrons and are 
thus also partons, but carry zero
baryon number. Their existence is indirectly confirmed by the observation of 
particle collisions from which 3 jets emerge, one of which must result
from a high-energy gluon formed in the initial collision,
subsequently decaying into a group of 
hadrons with net $B=0$.

\begin{figure}[htb]
\begin{center}
\epsfig{file=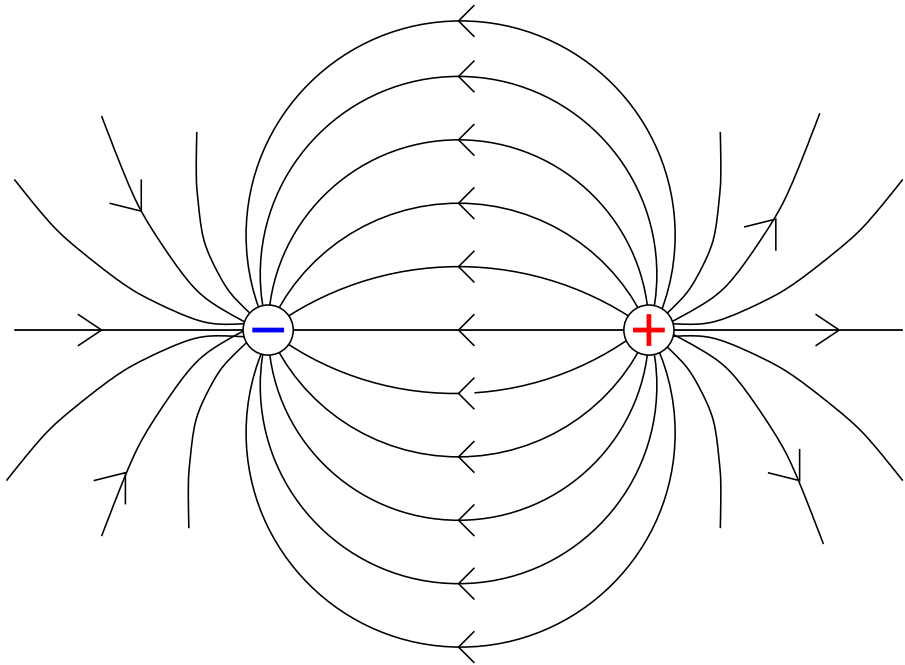, width=12cm}
\caption{Lines of electric flux 
between a pair of equal and opposite electric charges.\label{fig:coulomb}}
\end{center}
\end{figure}

The prototype for the quark-gluon interaction
is provided by a theory called Quantum Electrodynamics (QED),
which describes the force 
between electrically-charged particles 
such as electrons in terms of exchange of massless neutral 
quanta of the electromagnetic
field, namely photons. It can be shown that the force between two 
charges due to the single photon exchange
process analogous to that shown in 
Fig.~\ref{fig:confine}$a$ is described by a potential
\begin{equation}
V(r)={A\over r},
\end{equation}
where $A$ is proportional to the product of the charges;
in other words, single photon exchange reproduces the Coulomb
potential, and the resulting lines 
of electric flux between equal and opposite charges
trace out the familiar dipole field pattern, shown in Fig.~\ref{fig:coulomb}. 
In QCD the corresponding quantity is called {\sl chromoelectric flux\/}, and 
the gluon is the quantum of the chromoelectric and chromomagnetic fields.
By contrast with QED, however, single gluon
exchange in QCD only gives 
an accurate description of the force between quarks 
at very small distances. At larger separation, things become much more
complicated because 
as well as interacting with 
$q$ and $\bar q$, gluons can interact with
themselves (as explained below; see also Fig.~\ref{fig:confine}$b$), 
in contrast to photons which are electrically neutral and hence do not 
self-interact.
The gluons' additional ``stickiness'' presents a theoretical challenge
which to date has had no completely satisfactory solution; indeed, most of our
quantitative knowledge about the quark-gluon interaction at distance scales
$\gapprox O$(0.5fm) comes from formulating the equations of QCD on a 
discrete mesh of
spacetime points and modelling quantum fluctuations of the $q$, $\bar q$
and $g$
fields by numerical simulation.
A measure of the difficulty of this problem is that such {\sl lattice gauge
theory\/} simulations require dedicated use of 
the world's most powerful computers \cite{jbk}. 
Here we content ourselves with summarising
the result: it turns out that the potential between a $q\bar q$ pair 
at separation $r$ is 
\begin{equation}
V(r)=-{A(r)\over r}+Kr.
\label{eq:spot}
\end{equation}

For small $r$ the first term in (\ref{eq:spot}) dominates, and describes an 
attractive Coulomb-like interaction. It is important to note, however, that the
coefficient $A$ itself has a mild scale-dependence 
due to quantum effects. Detailed
analysis reveals that $A(r)\propto1/\ln(r^{-1})$, implying that 
the interaction between quarks gets weaker as
their separation decreases. In the limit $r\to0$ the quarks can be considered
non-interacting, a property known as {\sl asymptotic freedom\/} \cite{AF}.
Asymptotic freedom enables inelastic electron-proton collisions
to be interpreted in terms of scattering of high-momentum virtual photons off
almost-free partons; its theoretical discovery 
thus played a pivotal role in establishing QCD as the theory of the strong
interaction. 

As $r$ increases the second term in (\ref{eq:spot}) takes over,
implying that the $q\bar q$ potential rises linearly with
separation. 
\begin{figure}[htb]
\begin{center}
\epsfig{file=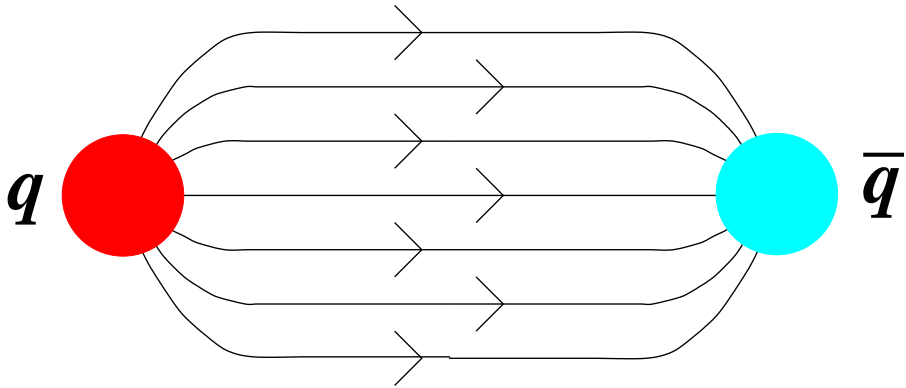, width=12cm}
\caption{Chromoelectric flux between a $q\bar q$ pair.\label{fig:fluxtube}}
\end{center}
\end{figure}
This can be understood by considering Fig~\ref{fig:fluxtube}, which
shows lines of chromoelectric flux between a quark and anti-quark. By contrast 
with QED, the field lines do not spread out in space to form
a dipole pattern, but remain localised within a 
narrow region of diameter $\sim0.7$fm 
between the sources known as a {\sl fluxtube\/} \cite{PGM}. Within the tube
the chromoelectric field strength is roughly uniform; therefore the energy 
of the system is proportional to the length of the fluxtube, in agreement
with eqn.(\ref{eq:spot}). In other words, the $q - \bar q$ 
force in this regime does not diminish with distance.
In this picture it is appropriate to consider hadrons
as little pieces of spinning chromoelectric ``string'', whose ends are defined
by the $q$, $\bar q$ moving at relativistic speeds. 
By matching the masses of the various observed hadrons
against their angular momentum (which of course 
is quantised in units of $\hbar$), it 
is possible to estimate the coefficient $K$ in (\ref{eq:spot}). Since it has
units of energy per unit length, it is known as the {\sl string tension\/}, 
with value $\simeq880\mbox{MeV/fm}\simeq(420\mbox{MeV})^2$,~\footnote{In
particle physics it is conventional to use units in which $\hbar=c=1$, in which
case energy, momentum and mass are all measured in the common unit MeV, and
distance and time, via the expression for the Compton wavelength 
$r=ct=\hbar/mc$, in units of (MeV)$^{-1}$. A useful conversion factor 
is 1fm$\simeq$(200MeV)$^{-1}$.} ie. a force sufficient to lift three
elephants!

At first sight eqn.(\ref{eq:spot}) suggests that 
the energy required to effect complete separation of the quarks,
that is, to ``ionise'' a hadron, is infinite. 
This is not strictly true,
because at some point the application of a 
non-relativistic concept such as $V(r)$ to relativistic quarks
must break down. Once the energy stored in the flux tube becomes comparable 
to twice the rest-energy of a quark, it becomes possible to break the string
by $q\bar q$ pair-production, the new particles 
acting as a fresh sink or source for 
the broken flux lines. The result is two $q \bar q$ mesons, each with a smaller
interquark separation than the original energetic meson. 
This process of particle production via
string-breaking helps to explain the 
profusion of hadrons formed in high-energy collisions.

Let us 
develop the analogy with QED a little further. The QCD quantity corresponding
to electric charge is called {\sl color\/}, and comes in three forms: ``red'',
``blue''. and ``green''. Both quarks and gluons have color quantum numbers. 
Anti-quarks have
complementary colors ``anti-red'' (shown as cyan in Fig.~\ref{fig:fluxtube}),
 ``anti-blue'' and ``anti-green''.
The gluon exchange responsible for QCD forces
carries color between different particles; hence the 
colored nature of the gluons can be regarded as the origin of
their self-interaction. 
As well as providing the theory with its name, 
the slightly frivolous
terminology
conveys in a heuristic way a deep dynamical principle of QCD.
Suppose we denote a green quark,
for instance, by $q_g$;
the only stable
finite-energy systems are those formed from complementary combinations
of color, such as $q_g\bar q_{\bar g}$ mesons or $q_r q_b q_g$ baryons.
The $q\bar q$ pair formed in string-breaking is produced with exactly the 
right color combination to maintain this overall color-neutrality.
Forces between color-neutral objects such as nucleons within the nucleus can be
viewed as a second-order effect akin to Van der Waals forces between neutral
atoms.\footnote{In fact inter-nucleon forces can be modelled by 
the exchange of color-neutral mesons such as pions. The dimension of a nucleus 
is comparable with the Compton wavelength of a pion.}
From our viewpoint the most important aspect is that colored objects 
such as isolated quarks or gluons are never observed. This property of QCD,
supported by all the theoretical considerations of the previous paragraphs,
and to date not contradicted by experiment, is called {\sl color
confinement\/}.

While a complete quantitative description of hadrons in QCD remains elusive, 
for many practical purposes including the thermodynamic arguments to be
developed in the next section, a simplified treatment
is both possible and desirable.
The two crucial 
ingredients we have identified, asymptotic freedom and color confinement, 
are built into a much simpler description of the strong interaction known as the
{\sl Bag model\/}, in which massless quarks  move freely within a
spherical hadron of radius $R$, but are prevented from travelling further by an
inwards-acting pressure due to the confining nature of the bulk vacuum
\cite{MITbag}. This
can be modelled by assigning a constant energy density $\Lambda_B^4$ 
to the non-confining
region within the hadron. 
The energy of a hadron of radius $R$ is then given by
\begin{equation}
E\sim R^3\Lambda_B^4+{C\over R},
\end{equation}
the second term arising from the kinetic energy of the confined 
quarks due to the
Uncertainty Principle. The hadron mass can be found by minimising $E$ 
with respect to $R$; eliminating the constant $C$  one finds
\begin{equation}
M\sim 4R^3\Lambda_B^4.
\end{equation}
Using typical values $M\sim1000$MeV, $R\sim1$fm, we derive a value for the
bag constant $\Lambda_B\sim200\mbox{MeV}$.

We now come to another important aspect of QCD dynamics. When a particle with 
spin $\vec s$ propagates, it is possible to define a quantity called {\sl
helicity\/} $h=\vec s.\vec k/\vert k\vert$, which is the projection
of the spin axis along the direction of the particle's motion, defined by the
momentum $\vec k$. For a spin-${1\over2}$ particle like the quark, there are
two possible helicity eigenstates $h=\pm{1\over2}$, 
usually referred to as left- and right-handed states since they 
are related by a mirror reflection. A quark's helicity is not altered by either
emission or absorption of a gluon;
hence in the absence of any other effect one
might deduce that the numbers of left- and right-handed quarks are separately
conserved in QCD, leading to two good quantum numbers $B_L$ and $B_R$.
A moment's thought, however, shows that this can only be the case if quarks
have zero mass and hence travel at the speed of light. Otherwise, it is 
possible to Lorentz boost to a frame in which the quark's momentum has the
opposite sign; since angular momentum along the
boost axis is not changed, helicity in the new frame must also have the opposite
sign. We conclude that in a relativistically covariant treatment massive 
quarks must be described as a superposition of helicity eigenstates, the mass
$m$ parametrising the overlap between them and hence effectively the rate
of $L\leftrightarrow R$ transitions. Since in this case
only $B=B_L+B_R$ remains as a good
quantum number, we say that the {\sl chiral symmetry\/} relating left and 
right-handed quarks and enabling them to be thought of as independent
particles 
is broken by the quark mass.

Chiral symmetry breaking ($\chi$SB), 
like that of other symmetries in many-body or quantum
field theory, can occur via the theory's own dynamics. We have seen 
that QCD is responsible for a strong attractive interaction between $q$ and
$\bar q$. The force is so strong, in fact, that the state usually considered
as the ground state or {\sl vacuum\/}, 
namely that of no particle present, is actually
unstable with respect to formation of a {\sl condensate\/} of tightly bound
$q\bar q$ pairs, much as the ground state of superfluid helium is a Bose
condensate of He atoms in the lowest quantum state. 
Let us denote the vacuum by the ket
$\vert0\rangle$, and the field operators which create or destroy a quark when 
acting on a ket
as $\bar\psi, \psi$ respectively. A $\chi$SB vacuum is then given by
\begin{equation}
\langle\bar\psi\psi\rangle\equiv\langle0\vert\bar\psi_L\psi_R
+\bar\psi_R\psi_L\vert0\rangle\not=0.
\label{eq:chisb}
\end{equation}
Since neither $\vert0\rangle$ is annihilated by $\psi$, nor
$\langle0\vert$ by $\bar\psi$, the vacuum must
contain $q\bar q$ pairs. In QCD it is believed that the value of
$\langle\bar\psi\psi\rangle\simeq(250\mbox{MeV})^3$, which can be interpreted
as the number of such pairs per unit volume.

As eqn.(\ref{eq:chisb}) implies, the condensate pairs $\psi_L$ with 
$\bar\psi_R$, and {\it vice versa\/}. 
Since $\bar\psi\psi$ leaves $B_L+B_R$ invariant but changes $B_L-B_R$ 
by two units, a non-vanishing condensate
implies that the latter quantity has no definite value in the vacuum and only
$B$ remains as a good quantum number.
A left-handed quark propagating through
such a vacuum can be annihilated by $\psi_L$, leaving $\bar\psi_R$
to create a right-handed quark with the same momemtum. The quark will thus 
flip its helicity at a rate proportional to $\langle\bar\psi\psi\rangle$ -- in
other words, it will propagate just as if it had a mass. 
We refer to this dynamically-generated mass as
the quark's {\sl constituent mass\/} $\Sigma$, 
as opposed to the quark's intrinsic
or {\sl current mass\/} $m$. If $\chi$SB occurs spontaneously by the formation
of a large $\langle\bar\psi\psi\rangle$, then $\Sigma$ may be very much greater
than $m$ due to this dynamical mass generation. 
We have listed estimates of both values for the known quarks
in Table~\ref{tab:quarks},
although since isolated quarks are not observed, neither quantity 
is unambiguously defined. If we look at the values for the $u$ and $d$ quarks,
however, we see that $m_{u,d}$ is not very much larger than
the electron mass 0.511MeV, whereas $\Sigma_{u,d}$ is 
comparable to one-third the nucleon mass 940MeV. 

Spontaneous $\chi$SB in 
QCD is
thus the agency by which the nucleon (and by extension the universe as a whole) 
acquires most of its mass.
It is also a natural consequence of quark
confinement \cite{Casher}. Consider the bag model description of massless 
quarks moving back
and forth within a small volume. At the surface of the bag the quarks must
reverse their direction of travel, 
but not their angular momentum, which is always
conserved. 
Therefore the interaction with the bag wall changes their helicity. Since this
cannot be achieved though any process involving gluon exchange between the
quarks in the bag, 
it must arise because the QCD vacuum in the volume outside the bag
contains a non-vanishing density of $q\bar q$ pairs with which the 
bag quarks can exchange helicity. Confinement implies $\chi$SB.

A useful analogy for $\chi$SB is supplied by 
the phenomenon of ferromagnetism in metals. On each atom of the metal,
occupying a site of a regular crystal lattice, there
is an unpaired electron carrying a magnetic moment or ``spin'' conveniently
denoted by $\uparrow$.
Quantum mechanical exchange
forces dictate that there is a tendency for spins on adjacent sites to 
align, although the preferred direction of alignment 
(ie. $\uparrow\uparrow$ vs. $\downarrow\downarrow$) is not determined by the 
microscopic Hamiltonian. However, if the temperature is low enough (ie.
$T<T_c$, the Curie temperature), then cooperative interactions amongst spins
at arbitrary separation result in a ground state in which 
a macroscopic fraction of the spins is aligned, resulting in the spontaneous
magnetisation of the sample $M=\langle\uparrow\rangle\not=0$
(angled brackets here denote a thermodynamic average rather than a 
quantum
expectation value).  
Since the magnetisation axis defines a particular direction 
in space, the original symmetry of the Hamiltonian under rotations of the spin
axis is spontaneously broken by $M\not=0$. The same effect can be promoted
by an external magnetic field $h\not=0$, which in this case explicitly breaks
the symmetry.
The relation between $M$ and $h$ exactly mirrors that between
$\langle\bar\psi\psi\rangle$ and $m$ in QCD.

It is interesting to examine the
spectrum of excitations above the ferromagnetic ground state $M\not=0$; 
it turns out that coherent
oscillations
of the spins in directions orthogonal to the 
magnetisation axis, known as {\sl spin-waves\/}, 
cost arbitrarily small amounts of energy to excite
in the limit of wavenumber $k\to0$. 
Spin-waves due to $\chi$SB in QCD 
correspond to massless bosonic particle excitations, which
are identified with the pion triplet $\pi^\pm,\pi^0$. With masses 135 - 140MeV,
these are by far the lightest hadrons, the next lightest meson 
made from $u,d$ quarks being the $\rho$
at 770MeV, and lightest baryon being the proton at 938MeV. The $\pi$
is not exactly massless due to the explicit $\chi$SB introduced by 
$m_{u,d}\not=0$.~\footnote{Arguments based on general features of chiral
symmetry breaking relate the pion mass to its size via $M_\pi\sim
R_\pi\sqrt{m_{u,d}\langle\bar\psi\psi\rangle}$.}
The $\pi$ plays a
special role in QCD -- its lightness provides the most direct evidence
for a vacuum with $\chi$SB, 
and many of its interactions were predicted purely on symmetry
grounds alone, long before the dynamics of QCD were worked out in any detail, or
the theory even formulated.

\section{The Quark-Gluon Plasma}
\label{sec:thermo}

A natural consequence of the composite nature of hadrons in QCD is that there
should be 
a phase transition as temperature $T$ is raised.
Let us start by
considering QCD with just $u$ and $d$ quarks, and ignoring the heavier flavors.
If there is no net
concentration of baryons (ie. $B=\mu=0$), then the dominant hadronic
degrees of freedom 
in a hot medium are initially pions $\pi^\pm,\pi^0$, 
which carry zero $B$ and can 
be relatively easily pair-produced. If we neglect their rest mass, which is
accurate for $T\gapprox100$MeV,~\footnote{We also use units in which
Boltzmann's constant $k_B=1$. A useful conversion is
$1\mbox{MeV}\simeq10^{10}$K.} then the
pressure due to a pion gas is given by adapting
the formula for blackbody radiation pressure:
\begin{equation}
P_\pi=-{{\partial\Omega_\pi}\over{\partial V}}\biggr\vert_{T,\mu}
=3\times\left({\pi^2\over90}\right)T^4,
\end{equation}
the factor three counting the number of pion charge states.
Now, the equivalent expression for a plasma of free light 
quarks and massless gluons which are no longer confined within hadrons
is much larger, because there are so many more degrees of freedom (ie. 
the plasma is a state of higher entropy):
\begin{equation}
P_{q\bar q}=2\times2\times3\times{7\over4}\times
\left({\pi^2\over90}\right)T^4\;\;\;;\;\;\;
P_g=2\times8\times\left({\pi^2\over90}\right)T^4.
\label{eq:SB}
\end{equation}
The numerical factors require explanation: for $q - \bar q$ pairs
there are 2 helicity
states each, 2 flavor states ($u$ and $d$), and 3 color states. 
The factor ${7\over4}$ arises due to the difference between Fermi-Dirac and
Bose-Einstein statistics. For $g$, there
are 2 helicity states and 8 color states (this is because gluons carry a 
combination of color and anti-color). Now, naively we expect the hadron gas and
quark-gluon plasma to be in equilibrium when the pressures coincide; however we
must also take confinement into account, 
most easily by considering the
bag constant of the previous section to act as a {\sl negative pressure\/}
for the
QGP \cite{bagT}. We arrive at
\begin{equation}
{1\over30}\pi^2T_c^4={37\over90}\pi^2T_c^4-\Lambda_B^4,
\end{equation}
whence the critical temperature for plasma formation, when 
the quarks are released from their confinement, is  $T_c\simeq144$MeV, 
somewhat over a trillion kelvin! 
The energy density of the plasma phase is predicted to be
$\varepsilon_{QGP}\simeq850\mbox{MeV/fm}^{3}$, and 
the latent heat at the transition 
$\Delta\varepsilon\simeq800\mbox{MeV/fm}^{3}$. 

\begin{figure}[htb]
\begin{center}
\epsfig{file=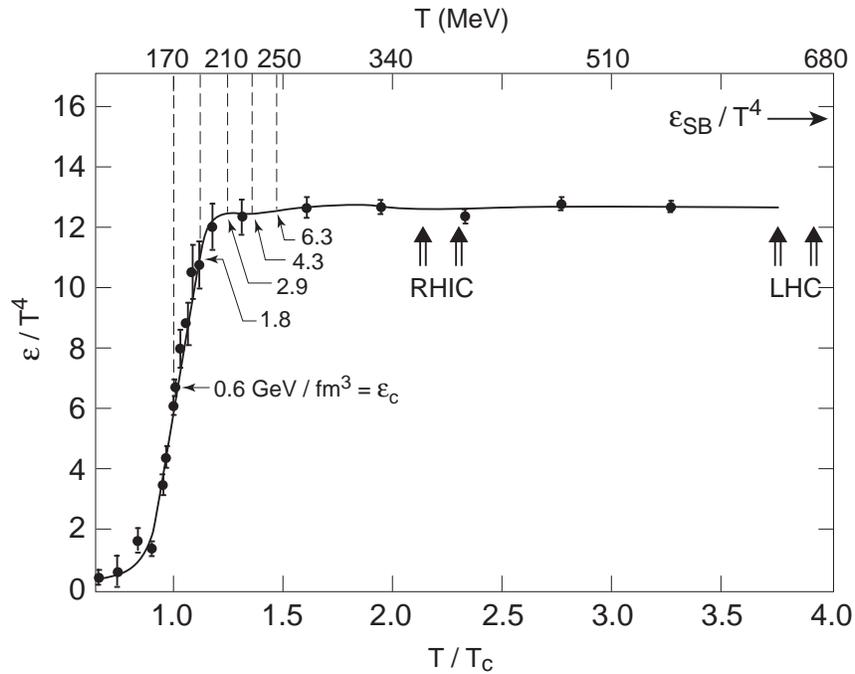, width=13cm}
\caption{Energy density $\varepsilon/T^4$ vs. temperature
$T/T_c$ for QCD with 3 light quark flavors. Heavy ion collisions at 
CERN's Super Proton Synchrotron 
probe energy densities in the neighbourhood of the ``knee'' of the curve, as 
described in Section~\ref{sec:rhic}.
RHIC and LHC refer to the 
regimes attainable at the Relativistic Heavy Ion and Large Hadron
Colliders.
(courtesy F. Karsch)
\label{fig:epsilon}}
\end{center}
\end{figure}
\begin{figure}[htb]
\begin{center}
\epsfig{file=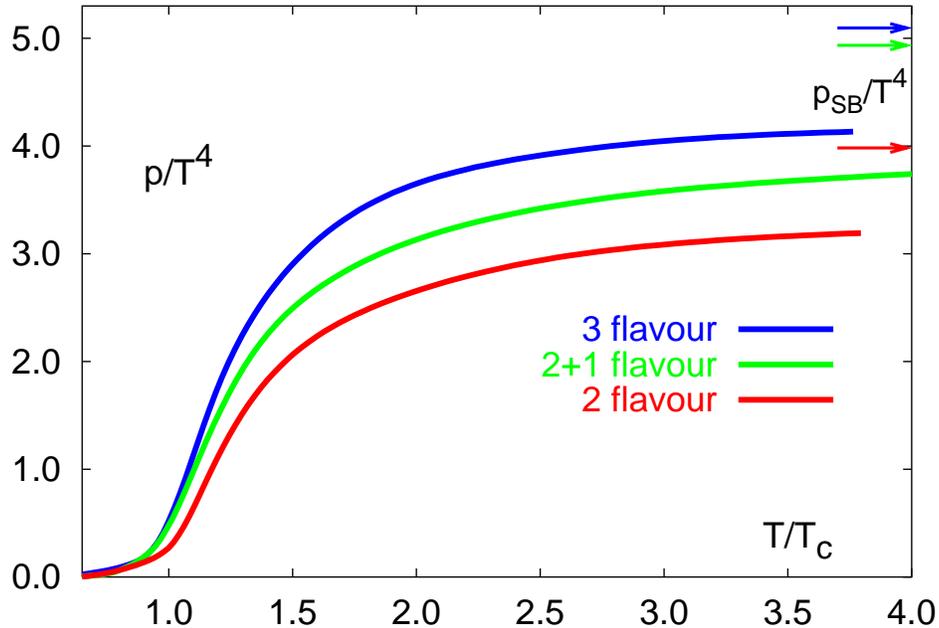, width=13cm}
\caption{Pressure $P/T^4$ vs. temperature $T/T_c$ for 2, 2+1 and 3 flavor QCD.
(courtesy F. Karsch)\label{fig:eos}}
\end{center}
\end{figure}
\begin{figure}[htb]
\begin{center}
\epsfig{file=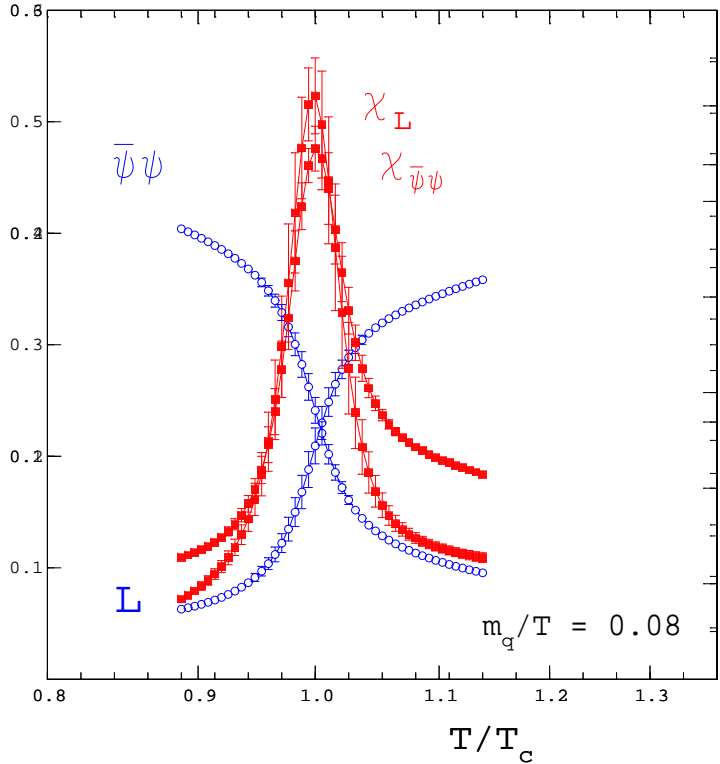, width=10cm}
\caption{Chiral condensate $\langle\bar\psi\psi\rangle$ and 
quark free energy function $L(f_q)$ as functions of $T$ in the 
neighbourhood of the transition (blue), together with their associated 
susceptibilities (red). (courtesy F. Karsch)\label{fig:poly}}
\end{center}
\end{figure}

The model we have used is crude, 
in effect treating the inside and outside of the bag
as different phases. More refined calculations of the transition to the QGP
are made using the lattice gauge theory techniques mentioned in the previous
section. In Figs.~\ref{fig:epsilon},~\ref{fig:eos} and \ref{fig:poly} 
we show the results of 
recent calculations performed on QCD with varying numbers of light quark
flavors \cite{Bielefeld}.
Fig.~\ref{fig:epsilon} shows the energy
density $\varepsilon$, 
and Fig.~\ref{fig:eos} the pressure $P$ as functions of $T$.
Each is expressed as a fraction of $T^4$, enabling comparison with the 
Stefan-Boltzmann predictions (\ref{eq:SB}) together with 
$\varepsilon_{SB}=3P_{SB}$, valid for a non-interacting relativistic plasma.
Fig.~\ref{fig:epsilon} shows the energy density rising very steeply at
$T\simeq170$MeV, and rapidly reaching a plateau at roughly 80\% of the SB
value. This is strong evidence for QGP formation at 
$T_c\simeq170\mbox{MeV}$, $\varepsilon_c\simeq600\mbox{MeV/fm}^{3}$, in rough
agreement with the bag model estimates. The disparity 
between the high-$T$ phase and a non-interacting
plasma, however, is reinforced by
Fig.~\ref{fig:eos}, which shows that 
for $T\lapprox3T_c$ the pressure falls appreciably below the SB
value. Strong interactions between $q$, $\bar q$ and $g$ persist in the QGP.
Only at very high temperatures when large energies are exchanged 
in inter-particle collisions
will the interaction strength weaken due to asymptotic freedom, 
making calculations 
using the single gluon exchange approximation feasible.

How else can the QGP be characterised? Fig.~\ref{fig:poly} 
shows the chiral condensate
$\langle\bar\psi\psi\rangle$ as a function of $T$, together with a quantity
$L$ related to the free energy $f_q(T)$ 
of an isolated quark via $L\propto\exp(-f_q/T)$. 
For $T<T_c$ $\langle\bar\psi\psi\rangle$ is large 
signalling 
$\chi$SB, and $L$ is small, signalling that the energy of an isolated color
source diverges, ie. confinement. Across the transition this behaviour
reverses itself indicating both 
that {\sl chiral symmetry is restored\/} 
and that {\sl color is no longer
confined\/} in the QGP. Also shown are the associated {\sl susceptibilities\/}
defined by $\chi_L=\langle L^2\rangle-\langle L\rangle^2$, and analogously
for $\chi_{\bar\psi\psi}$. These quantities indicate how strongly
thermodynamic observables fluctuate, and, as second derivatives of the free
energy, peak or even diverge at a phase transition.
It is a
remarkable fact that these
two {\it a priori\/} distinct transitions appear to coincide, as revealed
by the merging of the peaks in Fig.~\ref{fig:poly}.

The dominant degrees of freedom in the QGP are thus light $q$, $\bar q$ and $g$,
which justifies in retrospect the bag model treatment used
previously.
In terms of the $q\bar q$ potential (\ref{eq:spot}), 
the QGP is characterised
by the vanishing of the string tension $K$; indeed,
$V$ now takes a different form:
\begin{equation}
V_{QGP}(r)=-{C\over r}\exp(-r/\lambda_D),
\label{eq:screened}
\end{equation}
where $\lambda_D(T)\propto1/T$ 
is the {\sl Debye screening length\/} due to the non-zero
density of naked color sources in the QGP. For $T\gapprox T_c$ $\lambda_D\sim
O(0.1\mbox{fm})$ \cite{Kajantie}.

The value found for $T_c$ justifies the neglect of the heavy quarks $c$, $b$ and
$t$ in QCD thermodynamics, since their equilibrium 
concentration remains heavily 
suppressed by the Boltzmann factor $\exp(-m_{c,b,t}/T)$ 
in the critical region. We can thus restrict
our attention to $u$, $d$ and $s$.
In the limit $m_{u,d}\to0$ 
$\langle\bar\psi\psi\rangle$ can be regarded as an {\sl order parameter\/}
which vanishes in the QGP, just like the magnetisation
$M$ of a ferromagnet which vanishes at the Curie point. 
We have thus labelled ``hadronic'' and QGP phases in Fig.~\ref{fig:qcdphase}
by the value of $\langle\bar\psi\psi\rangle$ (shown in red). 
Another legitimate question is
the order of the phase transition;
the large latent heat predicted by the bag model indicates a strong
first order transition. However, our simple treatment has ignored any
possible variation with temperature of the bag constant $\Lambda_B$ itself.
In fact, lattice calculations with 3 light quark flavors
reveal $\Delta\varepsilon$ to be much smaller than the bag prediction
and hence the 
first-order transition much weaker; $\Delta\varepsilon$ may even vanish
altogether if $m_s$ is sufficiently large, ie. the 
transition is actually second order for just two light flavors \cite{order}. 
Now, since 
$m_{u,d}\not=0$
the order parameter $\langle\bar\psi\psi\rangle$ strictly never 
vanishes, but rather drops steeply in the transition region. For this reason
we have indicated QGP formation along the $\mu=0$ axis
in Fig.~\ref{fig:qcdphase} as a 
crossover rather than a true phase transition, with a possible critical point
ending a line of first-order transitions somewhere within the $(T,\mu)$ plane.
It should be stressed, however, that further numerical work is needed 
before this
picture can become conclusive.

\section{Relativistic Heavy-Ion Collisions}
\label{sec:rhic}

Having discussed the reasons for supposing that a new state of
strongly-interacting matter exists, we should now consider 
how it might be observed, which translates into the question of where to find or
how to produce temperatures of $O(10^{12}$K). A natural place to look is 
immediately after the Big Bang, when the energy density in the early universe 
considerably exceeded any found naturally today. 
In the first moments, energies were so
high that all matter was highly relativistic; under these circumstances
integration of the Friedmann equations governing the universe's expansion
predicts that the radius $R$ of the observable universe is related to 
the time after the Big Bang $t$ and temperature $T$ via
$R\propto T^{-1}\propto t^{1\over2}$; 
at later times when matter cooled and became
non-relativistic the behaviour changed over to $R\propto t^{2\over3}$.
We can use this information to extrapolate back from present conditions 
to predict that the QGP last existed between $10^{-5}$ and $10^{-4}$s after the
Big Bang. This is beyond the range of direct
observation, which cannot penetrate beyond
the epoch when the cosmic microwave background radiation was formed at roughly
$t\sim10^5$ years. The transition from quarks to hadrons may have left a
footprint, however, if it were first order. In this case, 
because of the non-zero interface
tension between the phases, the transition would have proceeded inhomogeneously,
eg. via the growth of bubbles of the hadronic phase at isolated points within 
the QGP. This would have resulted in local fluctuations in baryon 
concentration, which possibly had a significant impact on the relative
abundances of light elements formed at the nucleosynthesis epoch at $t\sim10$
minutes \cite{BBNS}.

In recent years, however, most attention has focussed on the possibility 
of recreating the QGP in terrestrial laboratories in 
{\sl relativistic heavy-ion collisions\/}, ie. high energy collisions between 
nuclei such as sulphur (S), lead (Pb) and gold (Au) \cite{heavy}. The first 
experiments have been
performed with fixed target nuclei at the 
Alternating Gradient Synchrotron (AGS) in Brookhaven and the 
Super Proton Synchrotron (SPS) at CERN,
with centre of mass (CM) energies of $2A$ and $18A$GeV respectively ($A$ is
the number of nucleons in the nucleus). Last year experiments with 
colliding nuclear beams commenced at 
RHIC in
Brookhaven, taking advantage of the higher energy of $200A$GeV available in the
CM frame; experiments are due 
to start at the ALICE
detector at the Large Hadron Collider (LHC) at
CERN in 2006, this time with a CM energy of $5500A$GeV. The regions of the
phase diagram probed by these experiments are indicated schematically 
by blue crosses in Fig.~\ref{fig:qcdphase}.
The collisions are incredibly
complex events, producing $O(10^3)$ charged particle tracks
at SPS and RHIC
energies 
and forecast to produce $O(10^4)$ at ALICE, and their study presents a
stringent challenge to data acquisition and storage techniques. An end-on view
of the charged tracks, ie. viewed along the beam, from a recent event 
observed at the STAR detector at RHIC
is shown in Fig.~\ref{fig:RHIC}.

\begin{figure}[htb]
\begin{center}
\epsfig{file=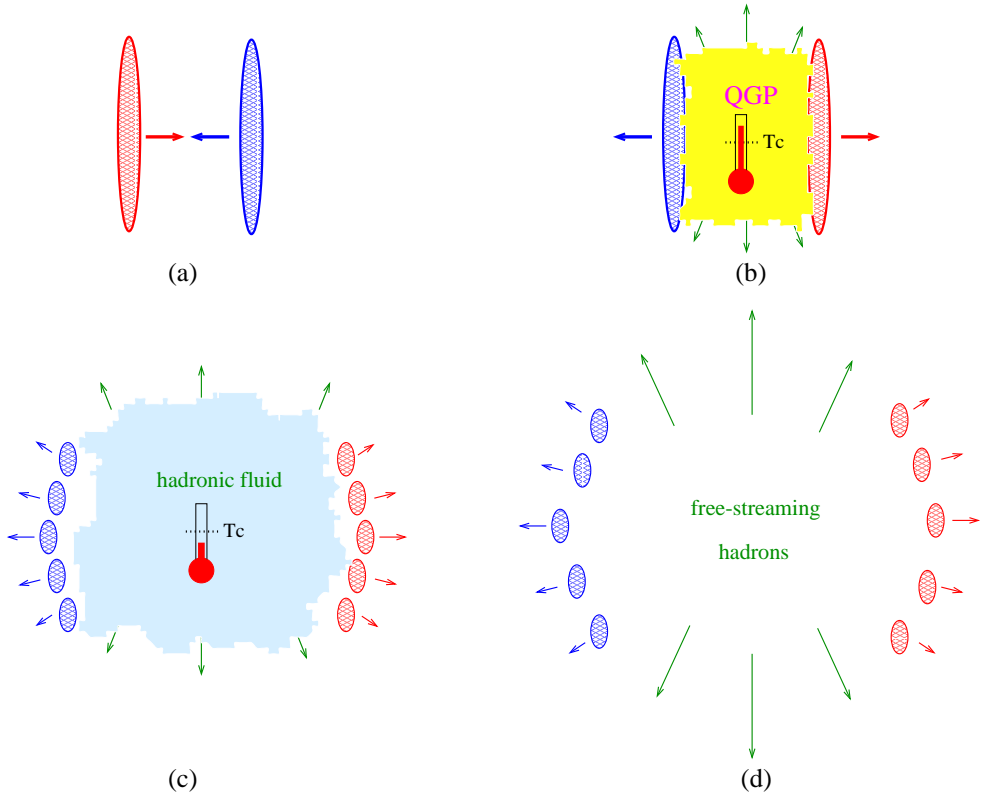, width=13cm}
\caption{Schematic view of the various stages of a heavy-ion collision.
The thermometers indicate when thermal equilibrium might be attained.
\label{fig:cartoon}}
\end{center}
\end{figure}
The various stages of a heavy-ion collision are portrayed in 
Fig.~\ref{fig:cartoon}. The colliding nuclei are envisaged as spherical in their
rest frame, but in the CM frame are Lorentz-contracted  along their direction 
of motion due to their relativistic velocity $\beta=v/c$: 
even at SPS energies the length contraction factor $\gamma(\beta)\simeq10$,
so the nuclei are best pictured as ``pancakes'' (see Fig.~\ref{fig:cartoon}$a$).
When the nuclei meet, the
initial events are high-energy inelastic collisions between individual nucleons,
in which many partons are liberated. 
Because the density of nuclear material is high, the
released partons have the opportunity to rescatter several times with the 
result that their momenta, initially highly correlated along the beam axis,
are redistributed 
and 
a substantial fraction of the incident kinetic energy is deposited in the CM
frame to produce a ``fireball'' in the {\sl mid-rapidity\/} region
$y\simeq0$.\footnote{ For a particle with 4-momentum $(E,\vec k)$,
the rapidity $y$ of 
a collision product is defined by 
$y={1\over2}\ln\left({{E+k_z}\over{E-k_z}}\right)$,
$z$ being the direction along the beam axis. Rapidity is a convenient 
variable in the study of relativistic collisions since under Lorentz boosts it 
simply changes by an additive constant.}
This energy is then available for conversion 
to hadrons via $q\bar q$ 
pair
production -- for this reason 
as many anti-baryons as baryons will be
produced in the mid-rapidity region, 
which thus has a net baryon density $n_B\simeq0$, 
whereas the forward and backward regions $\vert y\vert\gapprox1$ (at SPS
energies \cite{stop}), shown in red and blue in Fig.~\ref{fig:cartoon},
are relatively rich in baryons 
corresponding to the initial concentrations in the nuclei which have in
effect ``passed through'' each other.

A question then presents itself: is the energy density $\varepsilon$ produced
in the mid-rapidity region sufficient, assuming thermalisation, to effect a
phase transition to the QGP? An important result to help assess this possibility
is due to Bjorken \cite{bjork}: for a collision 
at the origin $(t,z)=(0,0)$ 
the energy density in the central region
at proper time $\tau=\sqrt{t^2-z^2}$ is given by
\begin{equation}
\varepsilon(\tau)=
{1\over{\tau{\cal A}}}{{dE_\bot}\over{dy}}\biggr\vert_{y=0},
\label{eq:bjorken}
\end{equation}
where ${\cal A}$ is the transverse area of the incident nuclei and
$dE_\bot/dy$ denotes the {\sl transverse} energy of the collision products (ie.
excluding kinetic energy due to motion parallel to $z$) per unit of
rapidity.\footnote{Note that the volume element
${\cal A}dz=\tau{\cal A}\cosh ydy$.}
For Pb+Pb collisions at SPS with zero impact parameter, we observe ${\cal
A}\simeq60\mbox{fm}^2$ and $dE_\bot/dy\simeq200$GeV at mid-rapidity 
\cite{heavy,dEdy}. 
Let us assume
that if thermalisation occurs it does so over a typical QCD timescale of
$\tau_0\sim1\mbox{fm}/c$ 
(the time taken for light to traverse a nucleon); eqn.(\ref{eq:bjorken})
then yields $\varepsilon\simeq3$GeV/fm$^3$. A comparison with
Fig.~\ref{fig:epsilon} suggests that a temperature $T\simeq200$MeV is 
reached, ie. the fireball is hot enough to form QGP, as shown in
Fig.~\ref{fig:cartoon}$b$.

If QGP is formed, it must quickly expand and cool due to its excess 
pressure with respect to the vacuum. At some point, therefore, $T$ falls
below $T_c$, and hadrons reform (Fig.~\ref{fig:cartoon}$c$). 
At around this point the composition of the 
hadrons formed, the majority of which are pions, is fixed -- this is  known as
{\sl chemical freezeout\/}. The resulting hadronic gas 
continues to cool until 
interaction rates become insufficient to maintain thermal equilibrium in
the expanding medium; this is
known as {\sl thermal freezeout\/}, from which point the hadrons are free to
stream away to be detected (Figs.~\ref{fig:RHIC},\ref{fig:cartoon}$d$). 
Information about conditions inside the fireball must be inferred
from hadrons emitted from the {\sl surface of last scattering\/} at thermal
freezeout, just as conditions in the early universe must be inferred from 
observations of the microwave background. 
Evidence for thermalisation comes from analysis of the distribution
of transverse mass $m_\bot=\sqrt{m_0^2+k_\bot^2}$, which is found to be 
approximately Boltzmann $\exp(-m_\bot/T)$ for a variety of different hadron
species.
Particularly useful information is
carried by pairs of identical 
particles such as $\pi^-\pi^-$. The wavefunction of such identical bosons
must be symmetric with respect to exchange of momenta; the
resulting correlation enables the size of the collision region to be inferred
from the momentum spectrum of the emitted particles \cite{heavy}. 
The principle is identical
to the two-photon {\sl intensity interferometry\/}
used to estimate stellar diameters in
astronomy \cite{HBT}. The combined results of these analyses
suggest a freezeout temperature 
$T\lapprox100$MeV, at which point the last scattering surface has radius
$\sim7$fm and expansion velocity
$\sim0.6c$ \cite{TWH}, all of which are consistent with the estimate of the 
initial energy density obtained using eqn.(\ref{eq:bjorken}).

\begin{figure}[htb]
\begin{center}
\epsfig{file=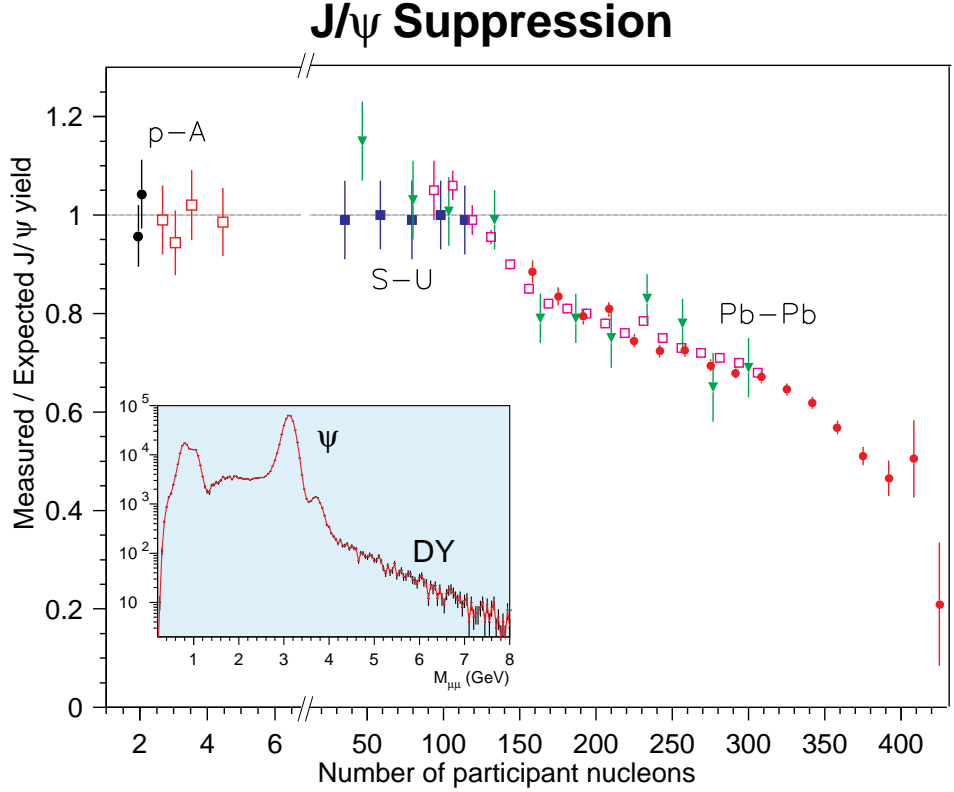, width=13cm}
\caption{$J/\psi$ suppression at the SPS. The data have been accumulated
from proton-nucleus as well as sulphur-uranium and lead-lead collisions. 
The inset shows the number of muon pairs produced in an ion collision as
a function of energy, clearly showing the peak due to $J/\psi$ decay 
at $\sim$3GeV, as well as a
high energy tail due to the Drell-Yan (single photon exchange) mode of
production. (courtesy C. Louren\c co)
\label{fig:jpsi}}
%\vskip 2 truecm
\end{center}
\end{figure}
Since it is plausible that thermalisation occurs in heavy-ion
collisions, and that the temperatures reached are of roughly the same order as
$T_c$, it is legitimate to consider possible signals for QGP formation. 
The first
is so-called {\sl $J/\psi$ suppression\/} \cite{MS}. 
Although charmed quarks are too heavy to be abundant 
in thermal equilibrium, they can be pair-produced in the initial
high-energy collisions to
form $c\bar c$ mesons, of which the $J/\psi$ at 3097MeV is the lightest --
in the vacuum it is relatively long-lived since it can only decay into lighter
hadrons via an intermediate pure-glue state which is hard to produce.
Now, 
bound states of heavy quarks $c,b,t$ are non-relativistic, and the
static potential $V(r)$ of eqn.(\ref{eq:spot}) is much more appropriate
for these systems than it is for mesons made from light quarks \cite{Eichten}. 
A consequence of QGP formation is that (\ref{eq:spot}) is replaced by the 
screened potential (\ref{eq:screened}):  the $c\bar c$ pair will thus
dissociate 
if the screening length $\lambda_D$ falls below the analogue of the Bohr radius
for the $J/\psi$.
If this happens 
charmed quarks will be far more likely to form mesons with open charm such 
as the $D$ ($c\bar u$, $c\bar d$), 
and hence fewer $J/\psi$ mesons should  be produced
in heavy-ion collisions as compared to similar energy $pp$ collisions.
The number of $J/\psi$'s produced can be estimated via the fraction
which decay not to hadrons
but to pairs of {\sl muons\/} $\mu^+\mu^-$. Muons are particles of mass 106MeV
which do not feel the strong interaction; though unstable, they survive
long enough to emerge from the fireball and be detected.
A compendium of recent results is shown in Fig.~\ref{fig:jpsi}
\cite{jpsidata}, where the 
suppression factor is plotted against the number of nucleons $N_{part}$
participating in the ion collision, 
which is proportional to the energy density
reached. Note that $J/\psi$ suppression is expected even in the absence of QGP,
due to dissociation on collision with other particles in a hadronic medium,
and can be observed in $pA$ collisions.
This effect can be modelled
by assuming a uniform decay rate integrated along the length of
nuclear material traversed by the $J/\psi$ before it emerges into the vacuum. 
The horizontal line in
Fig.~\ref{fig:jpsi} is normalised to this model expectation; there
is anomalous suppression for $N_{part}\geq130$, which is strong evidence for
quark-gluon plasma formation.

\begin{figure}[htb]
\begin{center}
\epsfig{file=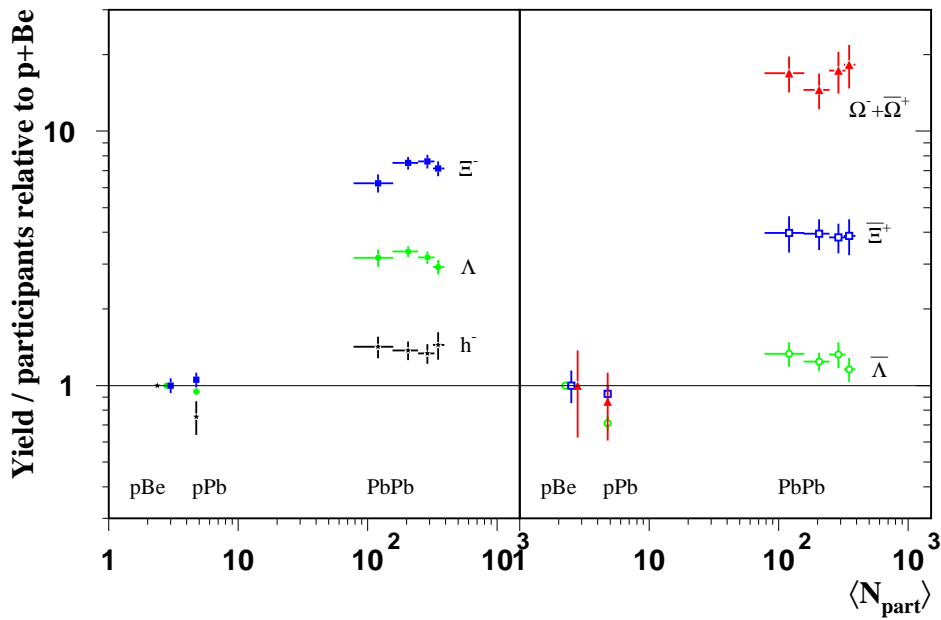, width=13cm}
\caption{Evidence for strangeness enhancement from the WA97 experiment at CERN.
(courtesy WA97 Collaboration)
\label{fig:wa97}}
\end{center}
\end{figure}
Another striking observation, {\sl strangeness enhancement\/},
comes from analysing the relative abundances of
hadron species in the collision products; it is found that many more hadrons 
containing $s$ and $\bar s$ quarks are produced 
in ion collisions than in similar energy $pp$ or
$pA$ collisions. Data from the WA97 experiment at the SPS (Fig.~\ref{fig:wa97})
show enhancement 
factors for $\Lambda$ ($uds$), $\Xi$ ($ssd$) and $\Omega$ ($sss$) baryons
and their corresponding anti-baryons,
showing that eg. $\Omega$ production is enhanced by a factor of 17 for ion
collisions with $N_{part}\geq10^2$ \cite{stenh}. What is the origin of this
effect? It has been proposed that $s\bar s$ pairs will be much easier to 
produce in the QGP than in a hadronic medium \cite{Raf}, 
both because they are considerably lighter due to chiral
symmetry restoration, and because of the much higher density of gluons, which 
opens up new formation processes such as gluon fusion $gg\to s\bar s$.
Note that the current mass $m_s$ is very similar in magnitude to 
$T_c$, which means that strangeness production should be a very
sensitive function of $T$ near the transition. Experiments currently under way
at the NA57 detector at SPS will explore intermediate values of
$N_{part}$, in the hope of finding a threshold for this effect. 

Heavy-ion collisions at the SPS have yielded signals
not easily explained in terms of purely hadronic physics. If the QGP has been
seen, however, 
its lifetime is probably $O$(1fm$/c$), comparable with the time taken for
its formation following thermal equilibration.
The current 
experiments at RHIC promise to produce QGP's at $T\simeq400$MeV, with 4 times
this lifetime, those 
at ALICE perhaps up to 10 times, making possible a detailed study of the QGP
independent of artifacts induced by the
proximity of the phase transition. Interpretation of $J/\psi$ suppression 
(and the
equivalent phenomenon in the $\Upsilon$ family
of $b\bar b$ systems),
and strangeness enhancement should become more clear-cut; additional
observations such as thermal photon emission (ie. blackbody radiation)
directly from the QGP will become feasible. Higher energies
will also create jets, and their interaction with and penetration of the 
plasma will yield much important information. Finally, improved detector
techniques will enable more detailed study of other thermal effects,
such as the downward shift and broadening of the $\rho$ meson resonance
\cite{ceres},
due respectively to partial chiral symmetry restoration and
collision-broadening\footnote{This is analagous to the pressure
broadening of spectral lines in atomic physics; the resonance width $\Gamma$
is related
to the $\rho$ lifetime $\Delta\tau$ via the Uncertainty Principle 
$\Gamma\sim1/\Delta\tau$, and hence
increases when the mean time between collisions is less than
the natural lifetime.}
in a hot hadronic medium.
In more ways than one, heavy-ion physics is entering
a golden age.

\section{Quark Matter and Color Superconductivity}

I now consider the behaviour of QCD as a function of chemical 
potential $\mu$. For $T$ strictly zero as $\mu$ increases, 
the ground state is initially the state
with no particle present, ie. the vacuum. This situation 
persists until $\mu$ reaches the value of the nucleon rest mass minus the
binding energy per nucleon in nuclear matter, when it becomes energetically
preferable to populate the ground state with a bound nucleon fluid. Ignoring
Coulomb repulsive forces between protons,\footnote
{Electromagnetic interactions, of
course, play a pivotal role in determining the size and stability of atomic 
nuclei, and
are also important in modelling neutron star interiors. On the energy 
scales depicted in Fig.~\ref{fig:qcdphase}, however, it is reasonable to
neglect them.}
this energy can be estimated
from nuclear physics as 16MeV/nucleon; therefore we can identify an
{\sl onset\/} value $\mu_o\simeq922$MeV at which point baryon density $n_B$
jumps from zero to nuclear density $n_{B0}\simeq0.16\mbox{fm}^{-3}$.
Since the vacuum and nuclear matter coexist at this point, the value $\mu=\mu_o$
corresponds to the ``room chemical potential'' that would be measured should
we ever be able to construct a suitable potentiometer!
Because $n_B=-{1\over V}{\partial\Omega\over\partial\mu}$, 
the discontinuity implies a 
first order phase transition. We expect the transition 
to persist for $T\not=0$ on grounds of continuity, and therefore show it 
as a coexistence line emerging from $\mu=\mu_o$ in
Fig.~\ref{fig:qcdphase}. As it now separates a phase in which baryons
can be present but are dilute from one in which they are condensed, it is 
known as the {\sl nuclear liquid-vapour transition\/} \cite{bagT,kap}.
It is anticipated that the line ends at a critical point with 
$T_c\simeq O(10\mbox{MeV})$; it is possible that critical opalescence has been
detected near this point in the form of the broad distribution of fragment 
sizes observed in medium-energy nuclear collisions \cite{Traut}.

What happens as $\mu$, and hence $n_B$, increase?
Unfortunately the lattice gauge theory simulations which were so useful
along the $T$-axis become ineffective once applied to QCD with $\mu\not=0$.
For densities up to 2 - 3$n_{B0}$ we can extrapolate from our current knowledge
of nuclear physics \cite{ST}. Beyond that we are forced to rely 
on approximate treatments
such as the bag model \cite{bagT}. As $n_B$ increases we again expect a 
transition from a 
phase in which matter exists in the form of nucleons to a higher entropy
phase where the dominant degrees of freedom are quarks. Naively this should
occur at densities 
of the order of a billion tonnes per teaspoonful
where the volume per baryon equals the baryon volume, and the bag surfaces
just touch.
For degenerate neutron matter at this critical density we have
\begin{equation}
n_{Bc}\simeq2\int_0^{k_F} {{4\pi k^2}\over{(2\pi)^3}}dk
={k_F^3\over{3\pi^2}}\simeq{{(\mu_c^2-M^2)}^{3\over2}\over
{3\pi^2}}\simeq{1\over R^3}\simeq1\mbox{fm}^{-3},
\end{equation}
giving $\mu_c\sim1200\mbox{MeV}$.
Various model estimates
yield $\mu_c\sim1100$ - 1500MeV, and the jump in density at the transition 
$\Delta n_B\sim2$ - $5n_{B0}$ \cite{Klev,HJSSV}. 

\begin{figure}[htb]
\begin{center}
\epsfig{file=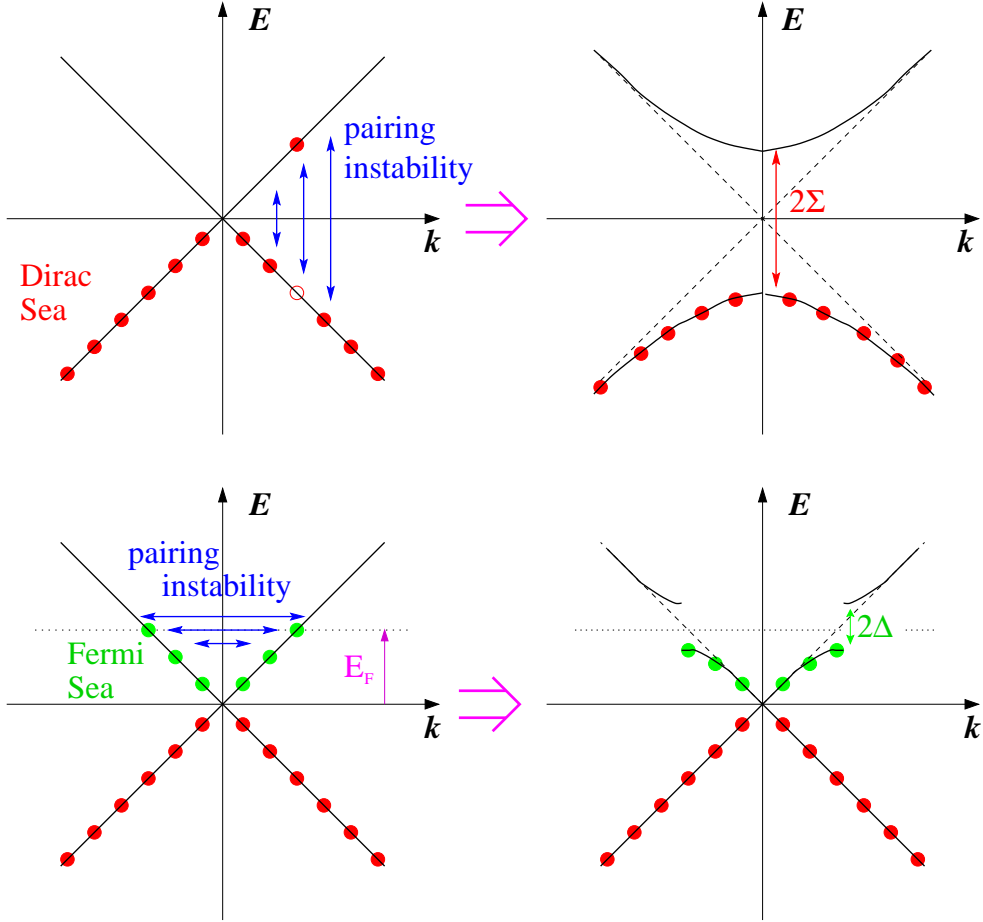, width=13cm}
\caption{Pairing instabilities leading to chiral symmetry breaking (top),
and superconductivity (bottom)\label{fig:disperse}}
\end{center}
\end{figure}
Let's discuss the nature of this 
{\sl quark matter\/} (QM) phase. 
In the bag model, the QM phase corresponds
to the bag interior in which chiral symmetry is restored and quarks are
light. That this might be so on more general grounds is illustrated in
the upper part of Fig.~\ref{fig:disperse}, where $\chi$SB is shown as due
to a pairing instability between quarks and anti-quarks of
equal and opposite momenta; the $\bar q$ are here interpreted 
as holes in the {\sl Dirac Sea\/} of negative energy states. $\chi$SB occurs 
if the binding energy of the $q\bar q$ pair exceeds the energy needed to
excite them: the result is the modified single-particle excitation spectrum 
shown at upper right, with an energy gap $2\Sigma$ between the highest occupied
and lowest empty states. Now when $n_B>0$, some positive energy 
states are also occupied, as shown at bottom left; we refer to these as 
belonging to the
{\sl Fermi Sea\/}. It is impossible to excite $q\bar q$ pairs if the $q$ state
has momentum $k<k_F$ and is hence already in the Fermi Sea; 
such pairs are {\sl Pauli-blocked\/}. At some
point, therefore, 
available $q\bar q$ states require so much energy to excite
that it is preferable to revert to a chirally symmetric ground state.
Since $k_F\gapprox\mu_c/3\gg m_{u,d}$, 
we deduce the quarks near the Fermi surface
which participate in QM's interaction with other forms of matter
are highly relativistic.

Where might QM be found? If the sequence of phases outlined above is
correct, then only at the onset $\mu=\mu_o$ does pressure $P=-\Omega/V$ vanish,
implying that nuclear matter at this point
is stable and self-bound. To reach the higher
densities needed for QM
an external pressure is needed; the most likely
source is the gravitational binding in the compact astrophysical objects
of mass
$O(10^{30}\mbox{kg})$ and radius $O(10\mbox{km})$ known as {\sl
neutron stars\/} \cite{ST}. Neutron star central densities are estimated as
lying in the range 5 - 10$n_{B0}$. It turns out that the 
maximum mass a neutron star can have before becoming unstable with respect to 
further gravitational collapse to form a black hole depends on how 
compressible its material is, as determined by the 
equation of state:
\begin{equation}
PV^\Gamma=\mbox{constant}.
\end{equation}
Roughly speaking \cite{ST}, $\Gamma\simeq{5\over3}$ for non-relativistic matter 
such as neutrons with $\mu\lapprox\mu_c$, 
producing a ``stiff'' equation of state and limiting masses
$\gapprox2M_\odot$, where $M_\odot$ is the mass of the sun; 
relativistic QM, on the
other hand, has $\Gamma\simeq{4\over3}$, 
implying a softer equation at the core, and
hence a lower limiting mass. So far observed neutron star masses have not 
exceeded 1.4$M_\odot$, which does not exclude models for their structure which
include a QM core \cite{henning}. It may well turn out that our best
experimental probe of QM will come through careful observation of the 
known neutron star population, currently $\sim10^3$, focussing on 
quantities such as the rate of change of angular momentum, cooling rate, and 
magnetic field.
One interesting possibility is that 
once $\mu/3\gapprox m_s$ the ground state of matter includes an appreciable
fraction of strange quarks, and hence a composition $\sim uds$; 
neutron stars may actually be made of such 
{\sl strange quark matter\/} (SQM) \cite{strange}. 
It is even conceivable, though
unlikely, that SQM is the ground state for $P=0$, and 
nuclear matter therefore only metastable.\footnote
{It has been suggested that catastrophic
consequences might follow if SQM were produced at rest 
in a heavy-ion collision at RHIC; the resulting ``strangelet'' might seed the
conversion of the entire Earth to SQM via weak interactions! Fortunately (at 
time of writing) this has not yet come about.}

Much recent interest has been aroused by the idea that QM might have
richer properties than those of a simple relativistic fermi liquid. 
Consider the lower panel of Fig.~\ref{fig:disperse}: if the 
$qq$ interaction is even weakly attractive at the Fermi surface, 
then another pairing instability, the
so-called {\sl BCS instability\/}, is expected between quark pairs at antipodal
points, leading to a ground state with a non-vanishing 
{\sl diquark condensate\/} $\langle qq\rangle\not=0$ \cite{BL,ARW,RSSV}.
Analogously to $\chi$SB, 
the instability leads to an energy gap $2\Delta$ between
highest occupied and lowest vacant one-particle states, 
the distinction being that this 
time the gap is located at the Fermi surface. In metals at temperatures of a 
few kelvin, 
a BCS instability can arise between {\sl Cooper pairs\/} of electrons
due to an 
attractive force arising from interaction with vibrations of the underlying 
crystal lattice of positively-charged ions. The Cooper pair condensate
leads to the phenomenon of {\sl superconductivity\/}, signalled by 
electric current flowing without resistance in a narrow layer close to the
sample surface;
the screening effect of this 
supercurrent results in the {\sl Meissner effect\/}, 
namely the complete exclusion
of magnetic field from the sample.
A BCS instability between neutral helium atoms in liquid $^3$He at milli-kelvin 
temperatures, on the other hand, leads to 
frictionless flow and quantisation of vorticity, 
a phenomenon known as {\sl superfluidity\/}. 

In QCD the force between two quarks due to single gluon exchange is
attractive (unlike single photon exchange between two electrons), implying
that a weak BCS instability should be present in QM \cite{BL}. More recent
calculations which attempt to model realistic strong interactions in the
regime $\mu\gapprox\mu_c$ predict a much bigger effect, with $\Delta$ 
as large as 10 - 100MeV \cite{BR}. 
What physical consequences
might arise? A crucial consideration, not applicable to $\chi$SB, 
is that the $qq$ wavefunction 
is constrained by the Pauli Exclusion Principle. As a result the ground state
is sensitive to the flavor composition of the available quarks.
Suppose $\mu/3$ is not much greater than $m_s$; in this case $k_{Fs}\ll
k_{Fu,d}$ and pairing is effectively restricted to the two light flavors.
The diquark condensate which thus forms is \cite{ARW,RSSV}
\begin{equation}
\langle qq\rangle_{2SC} = \epsilon^{\alpha\beta3}\epsilon_{ab}
\langle\psi^\alpha_a(k,\uparrow)
\psi^\beta_b(-k,\downarrow)\rangle\not=0.
\label{eq:2SC}
\end{equation}
The quark spins are combined in an antisymmetric singlet state; the overall
antisymmetry of (\ref{eq:2SC}) under quark exchange is then enforced by the 
alternating $\epsilon$ tensors acting on flavor $a,b=1,2$ and color
$\alpha,\beta=1,\ldots,3$ indices. Now, since the $qq_{2SC}$ wavefunction is not
color-neutral, we infer by analogy with the electrically-charged Cooper pair
that the ground state is {\sl color superconducting\/}. The most immediate
consequence is that out of the eight gluons, the five 
which carry color \#3
acquire a mass of $O(\Delta)$
and hence cannot penetrate QM over distances much greater than 
a screening length $\sim\Delta^{-1}$, 
in direct analogy with the Meissner effect 
in metallic superconductors.\footnote{In weak interaction physics 
the identical effect, 
now known as the {\sl Higgs phenomenon\/}, arises as a consequence of the 
condensation of a scalar {\sl Higgs field\/} and 
gives $W^\pm$ and $Z^0$ bosons masses of $O(100\mbox{GeV})$, restricting
the effective range of the weak interaction to $10^{-18}$m.} The 
three gluons left massless carry combinations of only the first two
colors. The $qq_{2SC}$
wavefunction however, like the QGP, respects chiral symmetry; for this
reason although the superconducting description will be the more
natural for $T\lapprox\Delta$, there may well be no true transition
separating 2SC and QGP phases, and we have shown them separated by a crossover
in Fig.~\ref{fig:qcdphase}.

For larger $\mu$, $k_{Fs}$ should increase up to the point where strange
quarks can participate in the pairing. In this case a more symmetric
``color-flavor-locked'' (CFL) condensate can form:
\begin{equation}
\langle qq\rangle_{CFL} = \sum_i\epsilon^{\alpha\beta i}\epsilon_{abi}
\langle\psi^\alpha_a(k,\uparrow)
\psi^\beta_b(-k,\downarrow)\rangle\not=0.
\label{eq:CFL}
\end{equation}
If anything the CFL state is still more exotic \cite{ARW}; all 8 gluons are
rendered massive implying color superconductivity, and chiral symmetry is
also broken.
Moreover, since the $qq_{CFL}$ 
operator either creates or annihilates two units of baryon number, 
$B$ is no longer a good quantum number -- it can be shown that 
this $B$-violation leads to superfluidity. The CFL state is currently believed
to be the stable ground state of matter as $\mu\to\infty$, and is shown as a
distinct phase in Fig.~\ref{fig:qcdphase}.

Observational and experimental evidence for these fascinating
new phases may be difficult to acquire. Since neutron star temperatures are 
typically $\lapprox10$MeV, 
it may seem that they are a natural place to look. However, two factors work
against us;
even if a neutron star has a QM core its bulk is 
probably normal nuclear matter, and color superconductivity is a phenomenon
confined to the vicinity of the Fermi surface rather than the whole Fermi
Sea and therefore has relatively little impact on the equation of state.
Both imply that any effect of color superconductivity is likely to be
quite subtle.
One prediction is that a color superconducting core would 
stabilise neutron star magnetic fields and prevent them from decaying over 
cosmologically significant timescales \cite{ABR}. Another speculation
concerns the formation
of a neutron star by core collapse of a very massive star during a
supernova explosion \cite{CR}. 
During this violent event the star cools from $\sim$50MeV to $\sim$10MeV 
over a timescale of a few
seconds by
emission of massless weakly-interacting particles called {\sl neutrinos\/} 
$\nu$,
$\bar\nu$.~\footnote
{An anti-neutrino is produced in the archetypal weak interaction,
$\beta$-decay of the neutron, via $d\to ue^-\bar\nu$.} 
If a transition to superconducting
QM occurs in this period, $\nu$ - $q$
scattering with $k\lapprox\Delta$ will be Pauli-blocked, with the 
effect of making the core effectively transparent to neutrinos.
They may emerge from the collapsing core in a sudden burst, rather
than steadily diffusing out over 10-20 seconds as in standard collapse
scenarios; it may be feasible to detect such a burst in terrestrial neutrino
detectors.

\section{Conclusion}

In this article I have explained how the study of the strong interaction 
has spawned a
new field, QCD thermodynamics, and argued that the resulting phase diagram
may be
surprisingly rich. An interesting feature for both $T>0$, $\mu>0$
has been the important role played by the strange quark; the fine
details of Fig.~\ref{fig:qcdphase} may be very sensitive to the precise
value of $m_s$. One aspect we should return to is the possibility of a 
second critical
point in the $(T,\mu)$ plane, whose existence 
follows if we assume that the thermal
transition is continuous or a crossover, whereas the density transition is 
first order \cite{Klev}. 
Its location is hard to calculate using
present technology (one estimate is $\sim(0.8T_c,0.6\mu_c)$ 
\cite{HJSSV}), but is expected to move to the left as $m_s$ decreases;
recall that for 3 light
quarks it should merge with the $T$-axis to produce a first-order thermal
transition. It has been proposed that 
various critical phenomena in the vicinity of this point could be detected in 
RHIC collisions \cite{SRS}. The reason that the strange quark has such a 
big influence on QCD thermodynamics is that $m_s$ is of the
same order of magnitude as the ``interesting'' QCD scale
200MeV$\simeq1\mbox{fm}^{-1}$ which has cropped up repeatedly in our discussion.
This scale characterises all QCD phenomena not described by single
gluon exchange;
there is a sense in which it is the scale at which the strong interaction
becomes ``strong''.

Apart from the obvious attraction of exploring new states of matter,
why is it important to continue study in this direction?
Firstly, QCD is an exceedingly challenging theory, and 
the experiments and observations I have sketched provide a new arena in which
to test its predictions. It is gratifying how far we can get via fairly
simple thermodynamic arguments, coupled with insights from relativity and
quantum theory. However, 
it is also a particularly satisfying area to work in for theorists
using systematic calculational techniques such as lattice
gauge theory, since by and large their theoretical predictions
still pre-date (and hence inform) experiment, 
rather than lag behind by, in certain cases, several tens of years.
Secondly, QCD thermodynamics deals with the only phase transition in
particle physics which will ever be studied under laboratory conditions.
Phase transitions in other particle physics contexts, 
such as electroweak or grand unified 
theories, are believed
to have played a crucial role in the first instants of the universe following
the Big Bang; they have been invoked in mechanisms to account for the
large-scale
structure of the universe as revealed by the distribution of 
galaxy clusters, the observed excess of matter over 
anti-matter, and also the period of early exponentially rapid 
expansion invoked to account for the observed homogeneity of the 
cosmic microwave background. A quantitative understanding of both 
equilibrium and non-equilibrium thermodynamics in these models will be required 
before the
full story can be told; QCD offers our sole opportunity of thoroughly
testing that
understanding.

%\newpage
\section*{Acknowledgements}
It is my pleasure to thank the Bielefeld lattice group led by Frithjof Karsch, 
the Birmingham ALICE group, and the authors of \cite{ARW}, not simply for their
help with preparing some of the figures, but also for introducing me to the 
complexities and excitement of this subject.

%\newpage
\section*{About the Author}
Simon Hands obtained a Ph.D. in theoretical particle physics from the University
of Edinburgh in 1986, and has subsequently held research posts at the
Universities of Oxford, Illinois and Glasgow, and in the Theory Division at 
CERN.
In 1992 he won a bottle of vintage champagne from the then Minister of Science,
William Waldegrave, for one of the five best responses to his challenge to
describe the Higgs boson on one A4 sheet! In 1993 he joined the 
newly-formed theory group at the University of Wales 
Swansea as a PPARC Advanced
Research Fellow, and is now Reader in Physics. Dr. Hands' research has
centred 
on using lattice gauge theory techniques to study
dynamical symmetry breaking in 
strongly-interacting quantum field theories like QCD, most recently 
specialising to the case of non-zero $\mu$ and $T$. Currently he enjoys the
use of a special-purpose multi-processor APEMille 
computer at Swansea, capable of over 20 billion
complex arithmetic operations per second, but more power is always welcomed.

\noindent
email: {\tt s.hands@swansea.ac.uk}


\begin{thebibliography}{xx}
%
\bibitem{quarks} M. Gell-Mann, Phys. Lett. {\bf8} (1964) 214;\\
G. Zweig, CERN reports TH-402, TH-412 (1964).
%
\bibitem{partons} R.P. Feynman, Phys. Rev. Lett. {\bf23} (1969) 1415;\\
J.D. Bjorken, Phys. Rev. {\bf179} (1969) 1547.
%
\bibitem{jbk} A very good introduction to the lattice method is
J.B. Kogut, Rev. Mod. Phys. {\bf55} (1983) 775. See also M. Creutz, 
{\sl Quarks, Gluons and Lattices\/}, (Cambridge University Press, 1983), and
I. Montvay and G. M\"unster,  {\sl Quantum Fields on a Lattice\/}, 
(Cambridge University Press, 1994). 
%
\bibitem{AF} D.J. Gross and F. Wilczek, Phys. Rev. Lett. {\bf30} (1973) 1343;\\
H.D. Politzer, Phys. Rev. Lett. {\bf30} (1973) 1346.
%
\bibitem{PGM} P. Pennanen, A.M. Green and C. Michael, Phys. Rev. {\bf D56}
(1997) 3903.
%
\bibitem{MITbag} A. Chodos, R.L. Jaffe, K. Johnson, C.B. Thorn and V.F.
Weisskopf, Phys. Rev. {\bf D9} (1974) 3471.
%
\bibitem{Casher} A. Casher, Phys. Lett. {\bf B83} (1979) 395.
%
\bibitem{bagT} see, eg.
J.I. Kapusta, {\sl Finite-Temperature Field Theory\/},
(Cambridge University Press, 1989).
%
\bibitem{Bielefeld} F. Karsch, E. Laermann and A. Peikert, Phys. Lett. 
{\bf B478} (2000) 447; Bielefeld report BI-TP2000/41 {\tt hep-lat/0012023},
and F. Karsch, private communication.  
%
\bibitem{Kajantie} K. Kajantie, M. Laine, J. Peisa, A. Rajantie, K. Rummukainen
and M. Shaposhnikov, Phys. Rev. Lett. {\bf79} (1997) 3130.
%
\bibitem{order} R.V. Gavai, J. Potvin and S. Sanielevici, Phys. Rev. Lett.
{\bf 58} (1987) 2519;\\
E. Laermann, Nucl. Phys. {\bf B}(Proc. Suppl.) {\bf63} (1998) 114.
%
\bibitem{BBNS} R.N. Boyd and T. Kajino, Ap. J. {\bf336} (1989) L55;\\
R.A. Maloney and W.A. Fowler, Ap.J. {\bf345} (1989) L5.
%
\bibitem{heavy} A good introduction is
C.-Y. Wong, {\sl Introduction to High-Energy Heavy-Ion Collisions\/}, 
(World Scientific, Singapore, 1994). A recent review of the experimental
situation at the SPS is U. Heinz, Nucl. Phys. {\bf A685} (2001) 414.
Also see {\tt http://cern.web.cern.ch/CERN/Announcements/2000/NewStateMatter}.
%
\bibitem{stop} NA49 Collaboration, H. Appelsh\"auser {\it et al\/},
Phys. Rev. Lett. {\bf82} (1999) 2471.
%
\bibitem{bjork} J.D. Bjorken, Phys. Rev. {\bf D27} (1983) 140.
%
\bibitem{dEdy} NA49 Collaboration, T. Alber {\it et al\/}, Phys. Rev. Lett.
{\bf75} (1995) 3814;\\
WA98 Collaboration, M. Aggarwal {\it et al\/}, Nucl. Phys. {\bf A610} (1996)
200c.
%
\bibitem{HBT} R. Hanbury-Brown and R.Q. Twiss, Nature {\bf127} (1956) 27.
%
\bibitem{TWH} B. Tom\'a\v sik, U.A. Wiedemann and U. Heinz, CERN report TH-215
(1999),
{\tt nucl-th/9907096}.
%
\bibitem{MS} T. Matsui and H. Satz, Phys. Lett. {\bf B178} (1986) 416;\\
H. Satz, Rep. Prog. Phys. {\bf63} (2000) 1511.
%
\bibitem{Eichten} E. Eichten, K. Gottfried, T. Kinoshita, K.D. Lane 
and T.M. Yan, Phys. Rev. {\bf D21} (1980) 203.
%
\bibitem{jpsidata} NA50 Collaboration, M. Abreu {\it et al\/}, Phys. Lett.
{\bf B477} (2000) 28.
%
\bibitem{stenh} WA97 Collaboration, E. Andersen {\it et al\/}, Phys. Lett.
{\bf B433} (1998) 209.
%
\bibitem{Raf} J. Rafelski and B. M\"uller, Phys. Rev. Lett. {\bf 48} (1982)
1066;\\
P. Koch, B. M\"uller and J. Rafelski, Phys. Rep. {\bf142} (1986) 167.
%
\bibitem{ceres} CERES Collaboration, B. Lenkeit {\it et al\/}, Nucl. Phys.
{\bf A661} (1999) 23c.
%
\bibitem{kap} L.P. Csernai and J.I. Kapusta, Phys. Rep. {\bf131} (1986) 223.
%
\bibitem{Traut} W. Trautmann, proceedings of {\sl 
International Summer School On Correlations And Clustering Phenomena In
Subatomic Physics\/} (Dronten 1996) p. 115, {\tt nucl-ex/9611002}.
%
\bibitem{ST} see, eg. S.L. Shapiro and S.A. Teukolsky, {\sl Black Holes, White
Dwarfs and Neutron Stars\/}, (Wiley, 1983).
%
\bibitem{Klev} S.P. Klevansky, Rev. Mod. Phys. {\bf 64} (1992) 649.
%
\bibitem{HJSSV}
M.A. Halasz, A.D. Jackson, R.E. Shrock, M.A. Stephanov and J.J.M. 
Verbaarschot, Phys. Rev. {\bf D58}:096007 (1998).
%
\bibitem{henning} A recent review is H. Heiselberg and M. Hjorth-Jensen,
Phys. Rep. {\bf328} (2000) 237.
%
\bibitem{strange} E. Witten, Phys. Rev. {\bf D30} (1984) 272;\\
E. Fahri and R.L. Jaffe, Phys. Rev. {\bf D30} (1984) 2397.
%
\bibitem{BL} D. Bailin and A. Love, Phys. Rep. {\bf107} (1984) 325.
%
\bibitem{ARW} M. Alford, K. Rajagopal and F. Wilczek, Phys. Lett. {\bf B422}
(1998) 247; Nucl. Phys. {\bf B537} (1999) 443.
%
\bibitem{RSSV}
R. Rapp, T. Sch\"afer, E.V. Shuryak and M. Velkovsky, Phys. Rev. Lett.
{\bf81} (1998) 53.
%
\bibitem{BR} J. Berges and K. Rajagopal, Nucl. Phys. {\bf B538} (1999) 215.
%
\bibitem{ABR} M. Alford, J. Berges and K. Rajagopal, Nucl. Phys. {\bf B571}
(2000) 269.
%
\bibitem{CR} G.W. Carter and S. Reddy, Phys. Rev. {\bf D62}:103002 (2000).
%
\bibitem{SRS} M.A. Stephanov, K. Rajagopal and E.V. Shuryak, Phys. Rev. Lett.
{\bf81} (1998) 4816.

\end{thebibliography}
\end{document}